\documentclass[aps,prd,reprint,superscriptaddress,longbibliography]{revtex4-2}

\usepackage{graphicx}
\usepackage{amsmath,amssymb, bm}
\usepackage{booktabs}
\usepackage{tabularx}
\usepackage{hyperref}

\hypersetup{
  colorlinks=true,
  linkcolor=blue,
  citecolor=blue,
  urlcolor=blue
}

\begin{document}

\title{Electromagnetic Field of an Oscillating Point Charge Measured by a Transversely Comoving Observer}

\author{Ruben Canora}
\affiliation{Elmore Family School of Electrical and Computer Engineering, Purdue University, West Lafayette, Indiana 47907, USA}

\author{Shengwang Du}
\email{dusw@purdue.edu}
\affiliation{Elmore Family School of Electrical and Computer Engineering, Purdue University, West Lafayette, Indiana 47907, USA}
\affiliation{Department of Physics and Astronomy, Purdue University, West Lafayette, Indiana 47907, USA}

\date{\today}

\begin{abstract}
We investigate the electromagnetic field produced by a harmonically oscillating point charge and measured by a spatially separated observer undergoing the same motion. We evaluate the exact retarded field on the observer's worldline and project the electromagnetic field tensor onto the observer's local orthonormal tetrad.  This construction yields the electric and magnetic fields, electromagnetic energy density, and Poynting vector measured as functions of the observer's proper time. Our formulation separates unambiguous local observables from the observer-dependent classification of a field as radiation and provides a concrete setting in which to examine how a charge that is spatially static in a non-inertial coordinate system can nevertheless produce time-dependent locally measured fields.
\end{abstract}

\maketitle

\section{Introduction}
\label{sec:introduction}

\begin{figure*}[t]
  \centering
  \includegraphics[width=0.9\textwidth]{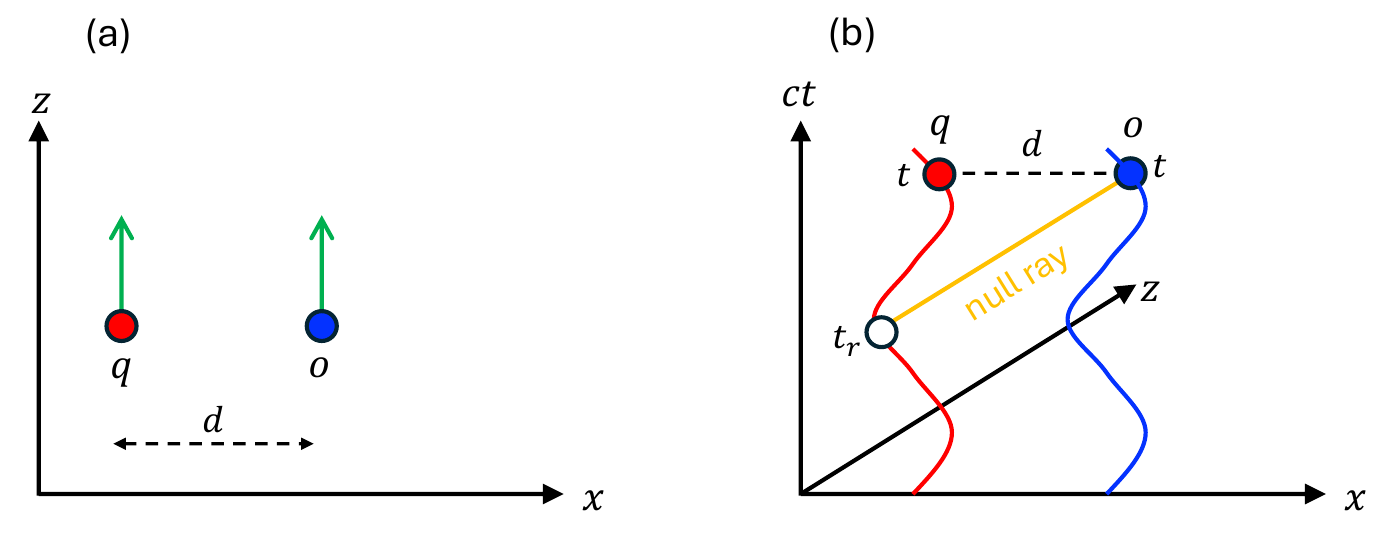}
  \caption{Source--observer configuration in an inertial reference frame $S$.  (a) A point charge $q$ and an
  observer $o$ comove along the $z$ direction while maintaining a constant
  transverse separation $d$ along $x$.  (b) Projected spacetime diagram of
  the two worldlines.  The field measured at the observation event
  $x_o^\mu(t)$ is determined by the null-separated source event
  $x_q^\mu(t_r)$.  The projection is schematic and not drawn to scale.}
  \label{fig:source-observer-configuration}
\end{figure*}

The electromagnetic field of an arbitrarily moving point charge is one of the fundamental exact solutions of classical electrodynamics.  In an inertial frame, the retarded Li\'enard--Wiechert solution separates naturally into a velocity field that decreases as the inverse square of the retarded distance and an acceleration field that decreases as its inverse \cite{Jackson1999,Rohrlich2007}.  For bounded, nonrelativistic motion, the latter produces the familiar far-zone radiation and Larmor power.  A subtler question arises when the measuring observer is itself accelerated and, in
particular, when the observer is instantaneously at rest relative to the charge.  Although the electromagnetic tensor at a specified event is covariant, its decomposition into electric and magnetic fields, its local energy flux, and the positive-frequency content relevant to a detector all depend on the observer's motion.  Moreover, radiation is often characterized through nonlocal information such as asymptotic energy flux or boundary conditions. The covariance of Maxwell's equations therefore does not by itself provide an observer-independent answer to the question of whether an
accelerated charge ``radiates'' according to a comoving observer.

This issue has a long history.  Born obtained an electromagnetic solution associated with hyperbolic motion in 1909 \cite{Born1909}.  Subsequent analyses revealed that the apparently simple question of whether a uniformly accelerated charge radiates depends critically on the spacetime region, observer congruence, boundary conditions, and adopted radiation criterion.  Fulton and Rohrlich argued that a uniformly accelerated charge has a radiative field relative to inertial observers \cite{FultonRohrlich1960}, while Rohrlich used the problem to clarify the local scope of the equivalence principle \cite{Rohrlich1963}.  Boulware showed that the radiative energy identified by inertial observers is associated with a region beyond the horizon accessible to uniformly coaccelerating observers \cite{Boulware1980}.  In the coaccelerating Rindler wedge, the field can be stationary and purely electric for the static observers.  This horizon-based resolution has been developed and explained further in Refs.~\cite{EriksenGron2004,AlmeidaSaa2006}.  These results demonstrate that the field measured by a specified observer and the flux assigned at an asymptotic boundary are distinct physical questions.

Several authors have developed radiation concepts intended for more general observer families.  Teitelboim formulated a covariant separation of the retarded electromagnetic energy--momentum into bound and radiative parts \cite{Teitelboim1970}.  Fugmann and Kretzschmar extended related ideas to non-inertial laboratories with general translational acceleration and rotation, emphasizing a suitably defined relative acceleration between the charge and the observer's frame \cite{FugmannKretzschmar1991}.  Leonov considered arbitrary point-charge motion and constructed special observer families for which the measured field is static and purely electric, subject to stated existence conditions \cite{Leonov2012}.  Conversely, operational thought experiments have been used to argue that an observer comoving with an accelerated charge may still register the mechanical effects of its field \cite{Broda2012}.  The variety of conclusions reflects, in part, the use of different observers and different meanings of detection and radiation.

The purpose of the present work is not to propose another universal definition of radiation.  Instead, we formulate a particular source--observer configuration for which the locally measured electromagnetic field is unambiguous.  In an inertial Minkowski frame, a point charge executes harmonic motion along the $z$ axis, while an observer undergoes the same motion at a fixed transverse displacement along $x$.  At every common inertial time, the source and observer therefore have identical velocities and accelerations.  The observer is neither a distant inertial observer nor one selected by imposing a static-field condition; in particular, it does not in general belong to the special observer family constructed in Ref.~\cite{Leonov2012}. 

The essential feature of this configuration is retardation.  The field at an observation event is determined by the source at an earlier, null-separated event.  Thus, equality of the simultaneous source and observer kinematics does not imply equality between the observer's state at the detection event and the source's state at the retarded event.  This retarded mismatch distinguishes the present periodic problem from the stationary Rindler description of eternal uniform proper acceleration.

Our calculation proceeds entirely from invariantly defined events and local measurements.  We first evaluate the retarded Li\'enard--Wiechert field tensor at the observer's event in the inertial frame.  We then project it, and the electromagnetic stress--energy tensor, onto a nonrotating orthonormal tetrad carried by the observer.  This gives the electric and magnetic fields, field energy density, and Poynting vector measured as functions of the observer's proper time.  The construction adopts the hypothesis of locality for an ideal pointlike accelerated observer \cite{Mashhoon1990}.  A detector with finite size, finite response time, or intrinsic memory would require an additional response model and lies outside the basic field calculation.

The remainder of this paper is organized as follows.  Section~\ref{sec:configuration} defines the worldlines, proper time, and kinematic quantities.  Section~\ref{sec:retarded-geometry} develops the exact retarded geometry.  Section~\ref{sec:inertial-field} specializes the Li\'enard--Wiechert solution to the source--observer configuration.  Section~\ref{sec:observer-measured-field} projects the electromagnetic tensor onto the observer's tetrad, and Sec.~\ref{sec:local-energy-flux} evaluates the locally measured energy density and Poynting vector.  Sections~\ref{sec:analytical-limits} and \ref{sec:proper-time-spectra} present analytical limits and proper-time spectra, respectively.  Section~\ref{sec:noninertial-coordinates} gives the complementary non-inertial coordinate description.  Section~\ref{sec:discussion} discusses the relation to previous observer constructions, the interpretation of the local energy flux, and the extent to which the measured field may be classified as radiation.  Section~\ref{sec:conclusions} summarizes the conclusions.  Technical conventions, component formulas, transport checks, numerical details, and asymptotic expansions are collected in the appendices.

\section{Physical configuration and conventions}
\label{sec:configuration}

A point charge $q$ moves in an inertial frame $S$ with Cartesian coordinates
$x^\mu=(ct,x,y,z)$ in Minkowski spacetime.  Greek indices run over
$0,1,2,3$, Latin spatial indices over $1,2,3$, and hatted indices label
components in an orthonormal tetrad.  We adopt the metric signature
$(-,+,+,+)$ and raise and lower inertial-coordinate indices with
$\eta_{\mu\nu}=\operatorname{diag}(-1,1,1,1)$
\cite{MisnerThorneWheeler1973}.  The self-field of the point charge is not
evaluated on its own worldline.

As illustrated in Fig.~\ref{fig:source-observer-configuration}(a), the charge moves harmonically along the $z$ axis with worldline
\begin{equation}
  x_q^\mu(t)
  =\bigl(ct,0,0,A\sin\omega t\bigr),
  \label{eq:charge-worldline}
\end{equation}
while an observer follows
\begin{equation}
  x_o^\mu(t)
  =\bigl(ct,d,0,A\sin\omega t\bigr),
  \label{eq:observer-worldline}
\end{equation}
where $A>0$ is the oscillation amplitude, $\omega>0$ is the angular frequency, and $d>0$ is the simultaneous transverse spatial separation along x direction in $S$.  Both worldlines are timelike provided
\begin{equation}
  \beta_0\equiv\frac{A\omega}{c}<1.
  \label{eq:beta-zero}
\end{equation}
The common coordinate velocity and acceleration are
\begin{align}
  \bm v_q(t)=\bm v_o(t)
    &=A\omega\cos(\omega t)\,\hat{\bm z},
    \label{eq:common-velocity}\\
  \bm a_q(t)=\bm a_o(t)
    &=-A\omega^2\sin(\omega t)\,\hat{\bm z}.
    \label{eq:common-acceleration}
\end{align}
Define
\begin{equation}
  \beta(t)=\beta_0\cos(\omega t),
  \qquad
  \gamma(t)=\frac{1}{\sqrt{1-\beta^2(t)}}.
  \label{eq:beta-gamma}
\end{equation}
Because the source and observer have the same velocity at equal coordinate time, their proper-time increments are identical,
\begin{equation}
  d\tau_q=d\tau_o=\frac{dt}{\gamma(t)}.
  \label{eq:proper-time-increment}
\end{equation}
Choosing $\tau_q=\tau_o=0$ at $t=0$, we may use a common proper-time function
\begin{equation}
  \tau(t)=\int_0^t\sqrt{1-\beta_0^2\cos^2(\omega t')}\,dt'.
  \label{eq:proper-time}
\end{equation}
This equality is a convenient synchronization inherited from $S$; it does not assert that spatially separated accelerated clocks define a global Einstein-synchronized non-inertial frame.

At an observation event $x_o^\mu(t)$, the field is determined by the charge at
an earlier, null-separated event $x_q^\mu(t_r)$.  The exact relation between
$t$ and the retarded time $t_r$ [Fig.~\ref{fig:source-observer-configuration}(b)], together with the resulting propagation
geometry, is developed in Sec.~III.

Finally, alongside $\beta_0$ we define the dimensionless separation
\begin{equation}
  \delta\equiv\frac{\omega d}{c}.
  \label{eq:delta}
\end{equation}
The pair $(\beta_0,\delta)$ will be used to display the relativistic and retardation dependence of all subsequent results.

\section{Exact retarded geometry}
\label{sec:retarded-geometry}

We now determine the source event that is causally connected to a specified
observation event, following the standard retarded construction
\cite{Jackson1999,Rohrlich2007}.  Let
\begin{equation}
  \begin{aligned}
  X^\mu(t,t_r)
  &\equiv x_o^\mu(t)-x_q^\mu(t_r) \\
  &=\bigl(c(t-t_r),d,0,
  A[\sin\omega t-\sin\omega t_r]\bigr)
  \end{aligned}
  \label{eq:null-separation-vector}
\end{equation}
be the separation four-vector from the source event to the observation event.
The retarded time is selected by the conditions
\begin{equation}
  X^\mu X_\mu=0,
  \qquad t_r<t,
  \label{eq:retarded-null-condition}
\end{equation}
which give
\begin{equation}
  c^2(t-t_r)^2
  =d^2+A^2\left[\sin(\omega t)-\sin(\omega t_r)\right]^2.
  \label{eq:retarded-time}
\end{equation}
Thus, the fixed simultaneous separation $d$ is not the source--observer
distance entering the retarded field.

It is useful to introduce the dimensionless phases and retardation interval
\begin{equation}
  \phi\equiv\omega t,
  \qquad
  \phi_r\equiv\omega t_r,
  \qquad
  \Delta\equiv\phi-\phi_r>0,
  \label{eq:dimensionless-phases}
\end{equation}
together with
\begin{equation}
  \chi(\phi,\phi_r)
  \equiv\sin\phi-\sin\phi_r.
  \label{eq:chi-definition}
\end{equation}
Using Eqs.~\eqref{eq:beta-zero} and \eqref{eq:delta}, the null condition becomes
the implicit dimensionless equation
\begin{equation}
  \Delta^2=\delta^2+\beta_0^2\chi^2.
  \label{eq:dimensionless-retarded-time}
\end{equation}
Equivalently, for a specified observation phase $\phi$, the physical root is
the positive solution of
\begin{equation}
  g(\Delta;\phi)
  \equiv \Delta^2-\delta^2
  -\beta_0^2\left[\sin\phi-\sin(\phi-\Delta)\right]^2=0.
  \label{eq:retarded-root-function}
\end{equation}

For completeness, the retarded solution exists and is unique throughout the
domain $d>0$ and $\beta_0<1$.  At fixed $t$, define
\begin{equation}
  f(t_r)=c(t-t_r)
  -\left|\bm r_o(t)-\bm r_q(t_r)\right|.
  \label{eq:retarded-existence-function}
\end{equation}
Because the source motion is bounded, $f(t_r)\to+\infty$ as
$t_r\to-\infty$, whereas $f(t)=-d<0$.  Continuity therefore guarantees at
least one root.  For every $t_r$,
\begin{equation}
  \frac{df}{dt_r}
  =-c+\bm n\cdot\bm v_q(t_r)
  =-c\left(1-\bm n\cdot\bm\beta_r\right)<0,
  \label{eq:retarded-uniqueness-derivative}
\end{equation}
where $\bm n$ is the unit vector from the source position at $t_r$ to the
fixed observation event, and $\bm\beta_r=\bm v_q(t_r)/c$.  The last inequality
follows from $|\bm\beta_r|<1$.  Hence $f$ is strictly decreasing, and the
intersection of the source worldline with the past light cone of
the observation event is unique.  The null condition also gives the useful
bounds
\begin{equation}
  \delta\leq\Delta\leq
  \sqrt{\delta^2+4\beta_0^2}.
  \label{eq:retardation-bounds}
\end{equation}

The spatial separation from the retarded source position to the observer is
\begin{equation}
  \bm R(t,t_r)
  \equiv\bm r_o(t)-\bm r_q(t_r)
  =d\,\hat{\bm x}+A\chi\,\hat{\bm z},
  \label{eq:retarded-spatial-vector}
\end{equation}
with magnitude
\begin{equation}
  R\equiv|\bm R|=c(t-t_r)=\frac{c}{\omega}\Delta.
  \label{eq:retarded-distance}
\end{equation}
The corresponding propagation direction is therefore
\begin{equation}
  \bm n\equiv\frac{\bm R}{R}
  =\frac{\delta}{\Delta}\,\hat{\bm x}
  +\frac{\beta_0\chi}{\Delta}\,\hat{\bm z},
  \qquad \bm n\cdot\bm n=1.
  \label{eq:retarded-unit-vector}
\end{equation}
The charge velocity and acceleration at the retarded event are
\begin{align}
  \bm\beta_r
  &\equiv\frac{\bm v_q(t_r)}{c}
  =\beta_0\cos\phi_r\,\hat{\bm z},
  \label{eq:retarded-beta}\\
  \dot{\bm\beta}_r
  &\equiv\left.\frac{d\bm\beta}{dt}\right|_{t_r}
  =-\beta_0\omega\sin\phi_r\,\hat{\bm z}.
  \label{eq:retarded-beta-dot}
\end{align}
Here and below, a subscript $r$ means that the source quantity is evaluated at
$t_r$.

The Li\'enard--Wiechert denominator is governed by
\begin{equation}
  \kappa_r
  \equiv1-\bm n\cdot\bm\beta_r
  =1-\frac{\beta_0^2\chi\cos\phi_r}{\Delta}>0.
  \label{eq:kappa-retarded}
\end{equation}
The three-vector retarded denominator is $R\kappa_r$.  In covariant notation,
if $u_r^\mu=\gamma_r(c,\bm v_q(t_r))$ is the source four-velocity, then
\begin{equation}
  -\frac{X_\mu u_r^\mu}{c}
  =\gamma_r R\kappa_r,
  \qquad
  \gamma_r=\frac{1}{\sqrt{1-\beta_0^2\cos^2\phi_r}}.
  \label{eq:covariant-retarded-distance}
\end{equation}
Equation~\eqref{eq:covariant-retarded-distance} distinguishes the invariant
retarded distance associated with the source four-velocity from the factor
$R\kappa_r$ appearing in the standard inertial-frame form of the
Li\'enard--Wiechert fields.

Differentiating the null condition along the observer's worldline gives the
rate at which the retarded time changes with the observation time.  Since
$c(t-t_r)=R$, one obtains
\begin{equation}
  c\left(1-\frac{dt_r}{dt}\right)
  =\bm n\cdot\left[
  \bm v_o(t)-\bm v_q(t_r)\frac{dt_r}{dt}\right].
  \label{eq:retarded-time-derivative-intermediate}
\end{equation}
Consequently,
\begin{equation}
  \frac{dt_r}{dt}
  =\frac{1-\bm n\cdot\bm\beta_o(t)}
  {1-\bm n\cdot\bm\beta_r}
  \equiv\frac{\kappa_o}{\kappa_r}>0,
  \label{eq:retarded-time-derivative}
\end{equation}
where
\begin{equation}
  \kappa_o
  =1-\frac{\beta_0^2\chi\cos\phi}{\Delta}.
  \label{eq:kappa-observer}
\end{equation}
Equivalently, $d\phi_r/d\phi=\kappa_o/\kappa_r$.  The positivity of both
factors shows explicitly that the retarded source phase advances monotonically
with the observation phase, as required by causality.

The periodic motion imposes useful exact symmetries.  Uniqueness of the
retarded root implies
\begin{align}
  \phi_r(\phi+2\pi)&=\phi_r(\phi)+2\pi,
  \label{eq:retarded-full-period-symmetry}\\
  \phi_r(\phi+\pi)&=\phi_r(\phi)+\pi.
  \label{eq:retarded-half-period-symmetry}
\end{align}
Thus, $\Delta(\phi)$ is $\pi$ periodic.  Under a half-cycle shift,
$\chi\to-\chi$, $n_x\to n_x$, $n_z\to-n_z$, and
$\bm\beta_r\to-\bm\beta_r$, while $R$ and $\kappa_r$ remain unchanged.  These
relations provide stringent checks on the analytical and numerical fields
derived below.

Several limits are also immediate.  For a static source, $\beta_0=0$, one has
\begin{equation}
  \Delta=\delta,
  \qquad t_r=t-\frac{d}{c},
  \qquad \bm n=\hat{\bm x},
  \qquad \kappa_r=1.
  \label{eq:static-retarded-limit}
\end{equation}
For fixed $\delta>0$ and $\beta_0\ll1$,
\begin{equation}
  \Delta
  =\delta+\frac{\beta_0^2}{2\delta}
  \left[\sin\phi-\sin(\phi-\delta)\right]^2
  +O(\beta_0^4).
  \label{eq:nonrelativistic-retardation-expansion}
\end{equation}
For $\delta\ll1$ at fixed $\beta_0<1$, expansion about the observation event
instead gives
\begin{equation}
  \Delta=\gamma(\phi)\,\delta+O(\delta^2),
  \label{eq:small-separation-retardation}
\end{equation}
where $\gamma(\phi)$ is defined in Eq.~\eqref{eq:beta-gamma}.  The field itself
is not evaluated at $d=0$, where the observer worldline coincides with the
point-charge worldline and the idealized self-field is singular.

For numerical calculations, Eq.~\eqref{eq:retarded-root-function} may be solved
with a bracketed one-dimensional root finder on the interval in
Eq.~\eqref{eq:retardation-bounds}.  A valid solution must satisfy
$\Delta>0$, the null residual $|g|$ must be below the chosen tolerance, and the
independent checks $\bm n^2=1$, $\kappa_r>0$, and
Eq.~\eqref{eq:retarded-time-derivative} should hold.  These checks will be used
when evaluating the exact Li\'enard--Wiechert field in Sec.~\ref{sec:inertial-field}.

\section{Retarded electromagnetic field in the inertial frame}
\label{sec:inertial-field}

We next evaluate the retarded electromagnetic field at the event
$x_o^\mu(t)$ in the inertial frame $S$.  All source quantities in this section
are evaluated at the unique retarded time $t_r$ determined in Sec.~\ref{sec:retarded-geometry}.  These
inertial-frame fields are evaluated on the accelerated observer's worldline,
but they are not yet the electric and magnetic fields measured in the
observer's local orthonormal frame.  We carry out that projection in Sec.~\ref{sec:observer-measured-field}.

In the inertial frame $S$, the exact electric field at $x_o^\mu(t)$ takes Li\'enard--Wiechert formulism \cite{Jackson1999,Rohrlich2007,LandauLifshitz1975}: 
\begin{equation}
  \bm E(t)
  =\frac{q}{4\pi\epsilon_0}
  \left[
  \frac{(1-\beta_r^2)(\bm n-\bm\beta_r)}
       {\kappa_r^3R^2}
  +\frac{\bm n\times
  \left[(\bm n-\bm\beta_r)\times\dot{\bm\beta}_r\right]}
       {c\kappa_r^3R}
  \right],
  \label{eq:lienard-wiechert-electric}
\end{equation}
and the magnetic field is
\begin{equation}
  \bm B(t)=\frac{1}{c}\bm n\times\bm E(t).
  \label{eq:lienard-wiechert-magnetic}
\end{equation}
Equations~\eqref{eq:lienard-wiechert-electric} and
\eqref{eq:lienard-wiechert-magnetic} are the standard retarded
Li\'enard--Wiechert fields \cite{Jackson1999,Rohrlich2007,LandauLifshitz1975}.
Here $\beta_r^2=\bm\beta_r\cdot\bm\beta_r$, and the overdot in
$\dot{\bm\beta}_r$ denotes differentiation with respect to the inertial source
time before evaluation at $t_r$.  The first term in
Eq.~\eqref{eq:lienard-wiechert-electric} decreases as $R^{-2}$ and will be
denoted the velocity field, whereas the second decreases as $R^{-1}$ and will
be denoted the acceleration field:
\begin{equation}
  \bm E=\bm E_{\mathrm{vel}}+\bm E_{\mathrm{acc}}.
  \label{eq:electric-field-split}
\end{equation}
This familiar split is useful in the inertial frame $S$ linked to the Coulomb contribution and radiation, but it is not by
itself an observer-independent decomposition into bound and radiative fields.

For the present geometry, define the retarded longitudinal velocity and
dimensionless acceleration coefficients
\begin{equation}
  b_r\equiv\beta_0\cos\phi_r,
  \qquad
  a_r\equiv\frac{\dot\beta_{r,z}}{\omega}
  =-\beta_0\sin\phi_r.
  \label{eq:retarded-b-a}
\end{equation}
Then
\begin{equation}
  \bm\beta_r=b_r\hat{\bm z},
  \qquad
  \dot{\bm\beta}_r=\omega a_r\hat{\bm z},
  \qquad
  \bm n=n_x\hat{\bm x}+n_z\hat{\bm z},
  \label{eq:planar-retarded-quantities}
\end{equation}
where
\begin{equation}
  n_x=\frac{\delta}{\Delta},
  \qquad
  n_z=\frac{\beta_0\chi}{\Delta},
  \qquad
  \kappa_r=1-b_r n_z.
  \label{eq:n-components-kappa}
\end{equation}
The vector triple product in the acceleration field reduces to
\begin{equation}
  \bm n\times
  \left[(\bm n-\bm\beta_r)\times\dot{\bm\beta}_r\right]
  =\omega a_r\left(n_xn_z\hat{\bm x}
  -n_x^2\hat{\bm z}\right).
  \label{eq:acceleration-triple-product}
\end{equation}
Therefore, the inertial-frame field at $x_o^\mu(t)$ has only three nonzero Cartesian
components,
\begin{equation}
  \bm E=E_x\hat{\bm x}+E_z\hat{\bm z},
  \qquad
  \bm B=B_y\hat{\bm y}.
  \label{eq:inertial-field-components}
\end{equation}
Separating the velocity and acceleration contributions gives
\begin{align}
  E_{x,\mathrm{vel}}
  &=\frac{q}{4\pi\epsilon_0}
  \frac{(1-b_r^2)n_x}{\kappa_r^3R^2},
  \label{eq:Ex-velocity}\\
  E_{z,\mathrm{vel}}
  &=\frac{q}{4\pi\epsilon_0}
  \frac{(1-b_r^2)(n_z-b_r)}{\kappa_r^3R^2},
  \label{eq:Ez-velocity}\\
  E_{x,\mathrm{acc}}
  &=\frac{q}{4\pi\epsilon_0}
  \frac{\omega a_r n_xn_z}{c\kappa_r^3R},
  \label{eq:Ex-acceleration}\\
  E_{z,\mathrm{acc}}
  &=-\frac{q}{4\pi\epsilon_0}
  \frac{\omega a_r n_x^2}{c\kappa_r^3R}.
  \label{eq:Ez-acceleration}
\end{align}
Using Eq.~\eqref{eq:lienard-wiechert-magnetic}, the magnetic components are
\begin{align}
  B_{y,\mathrm{vel}}
  &=\frac{q}{4\pi\epsilon_0 c}
  \frac{(1-b_r^2)n_xb_r}{\kappa_r^3R^2},
  \label{eq:By-velocity}\\
  B_{y,\mathrm{acc}}
  &=\frac{q}{4\pi\epsilon_0 c}
  \frac{\omega a_r n_x}{c\kappa_r^3R},
  \label{eq:By-acceleration}
\end{align}
with $E_x=E_{x,\mathrm{vel}}+E_{x,\mathrm{acc}}$,
$E_z=E_{z,\mathrm{vel}}+E_{z,\mathrm{acc}}$, and
$B_y=B_{y,\mathrm{vel}}+B_{y,\mathrm{acc}}$.  Although the source and observer
have equal velocities at the same inertial time, the generally nonzero $B_y$
depends on the source velocity and acceleration at $t_r$.  Equal-time
comoving motion therefore does not make the inertial-frame magnetic field
vanish.

For analytical and numerical work, it is convenient to introduce the signed
field normalization
\begin{equation}
  E_\omega\equiv\frac{q\omega^2}{4\pi\epsilon_0c^2}.
  \label{eq:field-normalization}
\end{equation}
Since $R=(c/\omega)\Delta$, the exact dimensionless components are
\begin{align}
  \frac{E_x}{E_\omega}
  &=\frac{(1-b_r^2)n_x}{\kappa_r^3\Delta^2}
  +\frac{a_r n_xn_z}{\kappa_r^3\Delta},
  \label{eq:dimensionless-Ex}\\
  \frac{E_z}{E_\omega}
  &=\frac{(1-b_r^2)(n_z-b_r)}{\kappa_r^3\Delta^2}
  -\frac{a_r n_x^2}{\kappa_r^3\Delta},
  \label{eq:dimensionless-Ez}\\
  \frac{cB_y}{E_\omega}
  &=\frac{(1-b_r^2)n_xb_r}{\kappa_r^3\Delta^2}
  +\frac{a_r n_x}{\kappa_r^3\Delta}.
  \label{eq:dimensionless-By}
\end{align}
Equations~\eqref{eq:dimensionless-Ex}--\eqref{eq:dimensionless-By}, together
with the unique solution of Eq.~\eqref{eq:dimensionless-retarded-time}, give
the complete exact field in $S$ as a function of $(\phi,\beta_0,\delta)$.

Several identities provide direct checks on these expressions.  First, the
acceleration field is transverse to the retarded propagation direction,
\begin{equation}
  \bm n\cdot\bm E_{\mathrm{acc}}=0.
  \label{eq:acceleration-field-transverse}
\end{equation}
The longitudinal part of the full electric field comes entirely from the
velocity field:
\begin{equation}
  \bm n\cdot\bm E
  =\frac{q}{4\pi\epsilon_0}
  \frac{1-b_r^2}{\kappa_r^2R^2}.
  \label{eq:longitudinal-electric-field}
\end{equation}
Because $\bm B=c^{-1}\bm n\times\bm E$, the two Lorentz invariants obey
\begin{align}
  \bm E\cdot\bm B&=0,
  \label{eq:field-invariant-pseudoscalar}\\
  E^2-c^2B^2
  &=\left(\bm n\cdot\bm E\right)^2
  =\left(\frac{q}{4\pi\epsilon_0}\right)^2
  \frac{(1-b_r^2)^2}{\kappa_r^4R^4}>0.
  \label{eq:field-invariant-scalar}
\end{align}
Thus the exact field at every finite nonsingular observation event is of
electric type.  Its leading far-zone acceleration field approaches a null
field, while the positive invariant in Eq.~\eqref{eq:field-invariant-scalar}
is supplied by the subleading velocity field.

The half-cycle symmetries derived in Sec.~III give
\begin{align}
  E_x(\phi+\pi)&=E_x(\phi),
  \label{eq:Ex-half-period}\\
  E_z(\phi+\pi)&=-E_z(\phi),
  \label{eq:Ez-half-period}\\
  B_y(\phi+\pi)&=-B_y(\phi).
  \label{eq:By-half-period}
\end{align}
In particular, $E_x$ contains only even harmonics of the mechanical frequency,
including a possible dc component, whereas $E_z$ and $B_y$ contain only odd
harmonics when expressed as functions of the inertial phase $\phi$.  These
selection rules will serve as useful spectral checks.

In the static limit $\beta_0\to0$, Eqs.~\eqref{eq:static-retarded-limit} and
\eqref{eq:dimensionless-Ex}--\eqref{eq:dimensionless-By} reduce to
\begin{equation}
  \bm E=\frac{q}{4\pi\epsilon_0d^2}\hat{\bm x},
  \qquad
  \bm B=0,
  \label{eq:static-coulomb-field}
\end{equation}
as required.  At large retarded distance, the velocity contribution decreases
as $R^{-2}$ and the acceleration contribution as $R^{-1}$.  This inertial-frame
scaling identifies the conventional far-zone field, but whether and how that
field is registered by the accelerated observer requires the local tetrad
projection developed next.

For later use, we adopt the field-tensor convention of
Refs.~\cite{Jackson1999,MisnerThorneWheeler1973}:
\begin{equation}
  F^{0i}=\frac{E_i}{c},
  \qquad
  F^{ij}=\epsilon^{ijk}B_k,
  \label{eq:field-tensor-convention}
\end{equation}
where $\epsilon^{123}=+1$.  This convention is consistent with the metric
signature $(-,+,+,+)$ and the Lorentz force law
$dp^\mu/d\tau=qF^{\mu\nu}u_\nu$.  For the component structure in
Eq.~\eqref{eq:inertial-field-components},
\begin{equation}
  F^{\mu\nu}=
  \begin{pmatrix}
    0 & E_x/c & 0 & E_z/c\\
    -E_x/c & 0 & 0 & -B_y\\
    0 & 0 & 0 & 0\\
    -E_z/c & B_y & 0 & 0
  \end{pmatrix}.
  \label{eq:inertial-field-tensor}
\end{equation}

\section{Electromagnetic field locally measured by the observer}
\label{sec:observer-measured-field}

The components obtained in Sec.~\ref{sec:inertial-field} refer to the inertial coordinate basis of
$S$.  To determine the field registered locally by the accelerated observer,
we project the electromagnetic tensor onto an orthonormal tetrad carried along
the observer's worldline.  We adopt the hypothesis of locality: at each event,
an ideal pointlike observer measures the same tensor components as an
instantaneously comoving inertial observer \cite{Mashhoon1990}.  This
prescription determines the local field uniquely once the observer's tetrad is
specified.

At the observation phase $\phi=\omega t$, define
\begin{equation}
  b_o\equiv\beta_0\cos\phi,
  \qquad
  \gamma_o\equiv\frac{1}{\sqrt{1-b_o^2}}.
  \label{eq:observer-beta-gamma}
\end{equation}
The observer's four-velocity is
\begin{equation}
  u_o^\mu=\frac{dx_o^\mu}{d\tau}
  =c\gamma_o(1,0,0,b_o).
  \label{eq:observer-four-velocity}
\end{equation}
A natural nonrotating tetrad adapted to this one-dimensional motion may be
constructed by a boost along $z$ \cite{MisnerThorneWheeler1973,MalufFaria2008}:
\begin{align}
  e_{\hat 0}{}^\mu&=\gamma_o(1,0,0,b_o),
  \label{eq:observer-tetrad-time}\\
  e_{\hat 1}{}^\mu&=(0,1,0,0),
  \label{eq:observer-tetrad-x}\\
  e_{\hat 2}{}^\mu&=(0,0,1,0),
  \label{eq:observer-tetrad-y}\\
  e_{\hat 3}{}^\mu&=\gamma_o(b_o,0,0,1).
  \label{eq:observer-tetrad-z}
\end{align}
Here $e_{\hat 0}{}^\mu=u_o^\mu/c$, while
$e_{\hat 1}{}^\mu$, $e_{\hat 2}{}^\mu$, and $e_{\hat 3}{}^\mu$ define the
observer's local $x$, $y$, and $z$ axes, respectively.  The orthogonality of the tetrade can be confirmed by the direct contraction $\eta_{\mu\nu}e_{\hat\alpha}{}^\mu e_{\hat\beta}{}^\nu=\eta_{\hat\alpha\hat\beta}$. This tetrad is also Fermi--Walker transported.  Let
\begin{equation}
  \alpha_o(t)\equiv\gamma_o^3 a_{o,z}(t)
  =-\gamma_o^3A\omega^2\sin\phi
  \label{eq:observer-proper-acceleration}
\end{equation}
be the signed proper acceleration along the local $z$ direction.  The
four-acceleration is
\begin{equation}
  a_o^\mu\equiv\frac{du_o^\mu}{d\tau}
  =\alpha_o e_{\hat 3}{}^\mu.
  \label{eq:observer-four-acceleration}
\end{equation}
The tetrad derivatives satisfy
\begin{align}
  \frac{de_{\hat 0}{}^\mu}{d\tau}
  &=\frac{\alpha_o}{c}e_{\hat 3}{}^\mu,
  &
  \frac{de_{\hat 3}{}^\mu}{d\tau}
  &=\frac{\alpha_o}{c}e_{\hat 0}{}^\mu,
  \label{eq:tetrad-derivatives}\\
  \frac{de_{\hat 1}{}^\mu}{d\tau}
  &=0,
  &
  \frac{de_{\hat 2}{}^\mu}{d\tau}
  &=0.
  \label{eq:transverse-tetrad-derivatives}
\end{align}
Equivalently, each tetrad vector obeys
\begin{equation}
  \frac{de_{\hat\alpha}{}^\mu}{d\tau}
  =\frac{1}{c^2}
  \left(u_o^\mu a_{o\nu}-a_o^\mu u_{o\nu}\right)
  e_{\hat\alpha}{}^\nu.
  \label{eq:fermi-walker-transport}
\end{equation}
There is therefore no rotation of the spatial triad relative to a local
gyroscope.  This property relies on the collinearity of the motion; no Thomas
rotation arises.

The tetrad components of the electromagnetic tensor are
\begin{equation}
  F_{\hat\alpha\hat\beta}
  =F_{\mu\nu}
  e_{\hat\alpha}{}^\mu e_{\hat\beta}{}^\nu.
  \label{eq:field-tensor-tetrad-projection}
\end{equation}
With the convention in Eq.~\eqref{eq:field-tensor-convention}, the locally
measured electric and magnetic components are defined by
\begin{equation}
  E_{\hat i}=-cF_{\hat 0\hat i},
  \qquad
  B_{\hat i}=\frac{1}{2}\epsilon_{ijk}F_{\hat j\hat k},
  \qquad \epsilon_{123}=+1.
  \label{eq:locally-measured-field-definition}
\end{equation}
Projection with Eqs.~\eqref{eq:observer-tetrad-time}--\eqref{eq:observer-tetrad-z}
is equivalent to a Lorentz boost of the inertial-frame field with velocity
$cb_o\hat{\bm z}$.  Using the planar field structure in
Eq.~\eqref{eq:inertial-field-components}, one obtains
\begin{align}
  E_{\hat x}
  &=\gamma_o\left(E_x-b_ocB_y\right),
  \label{eq:local-Ex-transformation}\\
  E_{\hat z}&=E_z,
  \label{eq:local-Ez-transformation}\\
  B_{\hat y}
  &=\gamma_o\left(B_y-\frac{b_o}{c}E_x\right),
  \label{eq:local-By-transformation}
\end{align}
with
\begin{equation}
  E_{\hat y}=B_{\hat x}=B_{\hat z}=0.
  \label{eq:vanishing-local-components}
\end{equation}
The absence of these components is an exact consequence of the planar
source--observer geometry, not an additional approximation. Substitution of Eqs.~\eqref{eq:dimensionless-Ex}--\eqref{eq:dimensionless-By}
gives closed expressions for the local observables.  In the normalization of
Eq.~\eqref{eq:field-normalization},
\begin{align}
  \frac{E_{\hat x}}{E_\omega}
  ={}&\frac{\gamma_o(1-b_r^2)n_x(1-b_ob_r)}
  {\kappa_r^3\Delta^2}
  +\frac{\gamma_o a_rn_x(n_z-b_o)}
  {\kappa_r^3\Delta},
  \label{eq:dimensionless-local-Ex}\\
  \frac{E_{\hat z}}{E_\omega}
  ={}&\frac{(1-b_r^2)(n_z-b_r)}
  {\kappa_r^3\Delta^2}
  -\frac{a_rn_x^2}{\kappa_r^3\Delta},
  \label{eq:dimensionless-local-Ez}\\
  \frac{cB_{\hat y}}{E_\omega}
  ={}&\frac{\gamma_o(1-b_r^2)n_x(b_r-b_o)}
  {\kappa_r^3\Delta^2}
  +\frac{\gamma_o a_rn_x(1-b_on_z)}
  {\kappa_r^3\Delta}.
  \label{eq:dimensionless-local-By}
\end{align}
These expressions, together with the retarded root
Eq.~\eqref{eq:dimensionless-retarded-time}, are the principal local-field
result.  They are exact for arbitrary $\beta_0<1$ and $\delta>0$.

Equations~\eqref{eq:dimensionless-local-Ex}--\eqref{eq:dimensionless-local-By}
also display why equal simultaneous motion does not generally eliminate the
locally measured magnetic field.  The boost is determined by $b_o$ at the
observation event, whereas the field contains the retarded quantities
$b_r=v_{q,z}(t_r)/c$ and $a_r=a_{q,z}(t_r)/(c\omega)$ from the earlier source
event.  In particular,
\begin{equation}
  B_{\hat y}=0
  \quad\Longleftrightarrow\quad
  cB_y=b_oE_x,
  \label{eq:local-magnetic-zero-condition}
\end{equation}
which is not an identity of the retarded solution.  The differences
$b_r-b_o$ and $n_z-b_o$, together with the acceleration-field term, encode the
retarded kinematic mismatch.

The local fields are functions of the observer's proper time.  From
Eq.~\eqref{eq:proper-time-increment},
\begin{equation}
  \frac{d\tau}{d\phi}
  =\frac{1}{\omega\gamma_o(\phi)},
  \qquad
  \tau(\phi)=\frac{1}{\omega}
  \int_0^\phi\sqrt{1-\beta_0^2\cos^2\phi'}\,d\phi'.
  \label{eq:proper-time-phase-map}
\end{equation}
Because $d\tau/d\phi>0$, this relation is invertible and defines
$\phi=\phi(\tau)$.  The proper-time period is
\begin{equation}
  T_\tau
  =\frac{1}{\omega}\int_0^{2\pi}
  \sqrt{1-\beta_0^2\cos^2\phi}\,d\phi
  =\frac{4}{\omega}E_{\mathrm{ell}}(\beta_0^2),
  \label{eq:proper-time-period}
\end{equation}
where
\begin{equation}
  E_{\mathrm{ell}}(m)
  \equiv\int_0^{\pi/2}\sqrt{1-m\sin^2\theta}\,d\theta
  \label{eq:elliptic-integral-second-kind}
\end{equation}
is the complete elliptic integral of the second kind.  The use of
$\sin^2\theta$ rather than $\cos^2\phi$ in this definition is immaterial under
a shift of the integration variable.

The half-cycle relations in Eqs.~\eqref{eq:Ex-half-period}--\eqref{eq:By-half-period},
together with $b_o(\phi+\pi)=-b_o(\phi)$ and
$\gamma_o(\phi+\pi)=\gamma_o(\phi)$, imply
\begin{align}
  E_{\hat x}\left(\tau+\frac{T_\tau}{2}\right)
  &=E_{\hat x}(\tau),
  \label{eq:local-Ex-half-period}\\
  E_{\hat z}\left(\tau+\frac{T_\tau}{2}\right)
  &=-E_{\hat z}(\tau),
  \label{eq:local-Ez-half-period}\\
  B_{\hat y}\left(\tau+\frac{T_\tau}{2}\right)
  &=-B_{\hat y}(\tau).
  \label{eq:local-By-half-period}
\end{align}
Thus $E_{\hat x}$ contains only even harmonics of the proper-time fundamental
frequency $2\pi/T_\tau$, including a possible dc component, whereas
$E_{\hat z}$ and $B_{\hat y}$ contain only odd harmonics.

Finally, the Lorentz invariants provide frame-independent checks on the tetrad
projection.  The local fields satisfy
\begin{align}
  \bm E_{\mathrm{loc}}\cdot\bm B_{\mathrm{loc}}&=0,
  \label{eq:local-field-invariant-pseudoscalar}\\
  E_{\mathrm{loc}}^2-c^2B_{\mathrm{loc}}^2
  &=E^2-c^2B^2
  =\left(\frac{q}{4\pi\epsilon_0}\right)^2
  \frac{(1-b_r^2)^2}{\kappa_r^4R^4}>0,
  \label{eq:local-field-invariant-scalar}
\end{align}
where
\begin{equation}
  \bm E_{\mathrm{loc}}
  =E_{\hat x}\hat{\bm e}_{\hat x}
  +E_{\hat z}\hat{\bm e}_{\hat z},
  \qquad
  \bm B_{\mathrm{loc}}
  =B_{\hat y}\hat{\bm e}_{\hat y}.
  \label{eq:local-field-vectors}
\end{equation}
These identities confirm that the tetrad projection changes the measured
electric and magnetic components without changing the invariant
electromagnetic field at the observation event.  A finite-size detector or a
detector with memory would require convolution of these local fields with a
specified response function; such nonlocal detector modeling is not included
here.

\section{Local energy density and energy flux}
\label{sec:local-energy-flux}

The electromagnetic stress--energy tensor provides the local energy
observables associated with a specified observer
\cite{Jackson1999,MisnerThorneWheeler1973}.  With the field-tensor convention
in Eq.~\eqref{eq:field-tensor-convention}, its inertial-coordinate form is
\begin{equation}
  T^{\mu\nu}
  =\frac{1}{\mu_0}
  \left(
  F^{\mu\alpha}F^\nu{}_{\alpha}
  -\frac{1}{4}\eta^{\mu\nu}
  F^{\alpha\beta}F_{\alpha\beta}
  \right).
  \label{eq:electromagnetic-stress-energy}
\end{equation}
Projection onto the observer's tetrad gives the energy density and energy flux
measured in the momentarily comoving orthonormal frame.  Equivalently, they may
be constructed directly from the local electric and magnetic fields obtained in Sec.~\ref{sec:observer-measured-field}.

The locally measured electromagnetic energy density is
\begin{equation}
  u_{\mathrm{loc}}
  =\frac{\epsilon_0}{2}
  \left(E_{\mathrm{loc}}^2+c^2B_{\mathrm{loc}}^2\right)
  =\frac{\epsilon_0}{2}
  \left(E_{\hat x}^2+E_{\hat z}^2+c^2B_{\hat y}^2\right).
  \label{eq:local-energy-density}
\end{equation}
The locally measured Poynting vector is
\begin{equation}
  \bm S_{\mathrm{loc}}
  =\frac{1}{\mu_0}\bm E_{\mathrm{loc}}
  \times\bm B_{\mathrm{loc}}.
  \label{eq:local-poynting-vector}
\end{equation}
For the exact planar field geometry,
\begin{align}
  S_{\hat x}
  &=-\frac{1}{\mu_0}E_{\hat z}B_{\hat y},
  \label{eq:local-Sx}\\
  S_{\hat y}&=0,
  \label{eq:local-Sy}\\
  S_{\hat z}
  &=\frac{1}{\mu_0}E_{\hat x}B_{\hat y}.
  \label{eq:local-Sz}
\end{align}
Thus a magnetic field along the local $y$ direction produces an energy flux in
the local $x$--$z$ plane.  The $x$ direction is the instantaneous transverse
direction from the source worldline toward the observer, whereas the $z$
direction is parallel to their common instantaneous motion.

It is useful to normalize these quantities by
\begin{equation}
  u_\omega\equiv\epsilon_0E_\omega^2,
  \qquad
  S_\omega\equiv\epsilon_0cE_\omega^2=cu_\omega,
  \label{eq:energy-flux-normalization}
\end{equation}
and to define
\begin{equation}
  e_{\hat x}\equiv\frac{E_{\hat x}}{E_\omega},
  \qquad
  e_{\hat z}\equiv\frac{E_{\hat z}}{E_\omega},
  \qquad
  b_{\hat y}\equiv\frac{cB_{\hat y}}{E_\omega}.
  \label{eq:dimensionless-local-fields}
\end{equation}
In terms of the exact dimensionless fields in
Eqs.~\eqref{eq:dimensionless-local-Ex}--\eqref{eq:dimensionless-local-By},
\begin{align}
  \frac{u_{\mathrm{loc}}}{u_\omega}
  &=\frac{1}{2}
  \left(e_{\hat x}^2+e_{\hat z}^2+b_{\hat y}^2\right),
  \label{eq:dimensionless-local-energy-density}\\
  \frac{S_{\hat x}}{S_\omega}
  &=-e_{\hat z}b_{\hat y},
  \label{eq:dimensionless-local-Sx}\\
  \frac{S_{\hat z}}{S_\omega}
  &=e_{\hat x}b_{\hat y}.
  \label{eq:dimensionless-local-Sz}
\end{align}
These equations give the complete local energy density and flux for arbitrary
$\beta_0<1$ and $\delta>0$ once the retarded root is known.

For any local observable $Q(\tau)$, define its average over one observer
proper-time period by
\begin{equation}
  \left\langle Q\right\rangle_\tau
  \equiv\frac{1}{T_\tau}\int_0^{T_\tau}Q(\tau)\,d\tau.
  \label{eq:proper-time-average-definition}
\end{equation}
Using Eq.~\eqref{eq:proper-time-phase-map}, the same average can be evaluated
over the inertial phase as
\begin{equation}
  \left\langle Q\right\rangle_\tau
  =\frac{1}{\omega T_\tau}
  \int_0^{2\pi}\frac{Q(\phi)}{\gamma_o(\phi)}\,d\phi.
  \label{eq:proper-time-average-phase}
\end{equation}
The factor $1/\gamma_o$ is essential: an unweighted average over the inertial
phase is not, in general, the average recorded by the observer's proper-time
clock.

The exact half-period relations in
Eqs.~\eqref{eq:local-Ex-half-period}--\eqref{eq:local-By-half-period} imply
\begin{align}
  u_{\mathrm{loc}}\left(\tau+\frac{T_\tau}{2}\right)
  &=u_{\mathrm{loc}}(\tau),
  \label{eq:energy-density-half-period}\\
  S_{\hat x}\left(\tau+\frac{T_\tau}{2}\right)
  &=S_{\hat x}(\tau),
  \label{eq:Sx-half-period}\\
  S_{\hat z}\left(\tau+\frac{T_\tau}{2}\right)
  &=-S_{\hat z}(\tau).
  \label{eq:Sz-half-period}
\end{align}
Consequently,
\begin{equation}
  \left\langle S_{\hat z}\right\rangle_\tau=0
  \label{eq:average-Sz-zero}
\end{equation}
exactly.  No corresponding symmetry forces
$\langle S_{\hat x}\rangle_\tau$ to vanish.  The transverse mean flux
\begin{equation}
  \left\langle S_{\hat x}\right\rangle_\tau
  =-\frac{1}{\mu_0T_\tau}
  \int_0^{T_\tau}E_{\hat z}(\tau)B_{\hat y}(\tau)\,d\tau
  \label{eq:average-Sx}
\end{equation}
must therefore be determined from the exact fields or from controlled
asymptotic expansions.  Its sign and magnitude may depend on
$(\beta_0,\delta)$.

The local energy flux is necessarily causal.  For arbitrary electromagnetic
fields,
\begin{equation}
  u_{\mathrm{loc}}^2-\frac{S_{\mathrm{loc}}^2}{c^2}
  =\frac{\epsilon_0^2}{4}
  \left(E_{\mathrm{loc}}^2-c^2B_{\mathrm{loc}}^2\right)^2
  +\epsilon_0^2c^2
  \left(\bm E_{\mathrm{loc}}\cdot\bm B_{\mathrm{loc}}\right)^2.
  \label{eq:energy-flux-causal-identity}
\end{equation}
Both terms on the right-hand side are nonnegative, and hence
\begin{equation}
  |\bm S_{\mathrm{loc}}|\leq cu_{\mathrm{loc}}.
  \label{eq:energy-flux-causal-bound}
\end{equation}
For the present exact field, Eq.~\eqref{eq:local-field-invariant-pseudoscalar}
sets the second term in Eq.~\eqref{eq:energy-flux-causal-identity} to zero,
whereas Eq.~\eqref{eq:local-field-invariant-scalar} makes the first term
strictly positive at every finite nonsingular observation event.  Therefore, $|\bm S_{\mathrm{loc}}|<cu_{\mathrm{loc}}$ at finite $R$.  Equality is approached only when the leading far-zone field
becomes null.

In the static limit, Eq.~\eqref{eq:static-coulomb-field} yields
\begin{equation}
  u_{\mathrm{loc}}
  =\frac{q^2}{32\pi^2\epsilon_0d^4},
  \qquad
  \bm S_{\mathrm{loc}}=0.
  \label{eq:static-local-energy-flux}
\end{equation}
In the far zone, the leading acceleration fields scale as $R^{-1}$, so their
energy density and flux scale as $R^{-2}$ and approach the null-field relation
$|\bm S_{\mathrm{loc}}|=cu_{\mathrm{loc}}$ at leading order.  The exact field
at finite distance nevertheless retains the positive invariant contribution
of the velocity field.

The quantities $u_{\mathrm{loc}}$ and $\bm S_{\mathrm{loc}}$ are unambiguous
pointwise observables for the specified observer.  A nonzero Poynting vector
means that this observer assigns a local electromagnetic energy flux to the
field at that event.  It does not, by itself, establish an observer-independent
classification as radiation, which generally requires additional nonlocal or
asymptotic information.  The analytical limits and numerical evaluation of
these observables will be developed in the following sections.

\section{Analytical limits}
\label{sec:analytical-limits}

The exact expressions derived above apply throughout the domain
$\beta_0<1$ and $\delta>0$.  We now examine several limits that clarify their
physical content and provide benchmarks for numerical calculations.  The
limits are taken at fixed observation phase unless stated otherwise.  Because
$E_\omega$ is independent of $d$ whereas the Coulomb field scales as $d^{-2}$,
dimensionless expressions may contain negative powers of $\delta$ even when
their physical interpretation is regular at fixed nonzero $d$.

\subsection{Nonrelativistic motion at arbitrary separation}
\label{subsec:nonrelativistic-limit}

Consider $\beta_0\ll1$ with fixed $\delta>0$.  Equation~\eqref{eq:nonrelativistic-retardation-expansion}
gives, to the order required here,
\begin{equation}
  \Delta=\delta+O(\beta_0^2),
  \qquad
  \phi_r=\phi-\delta+O(\beta_0^2).
  \label{eq:nonrelativistic-retarded-phase}
\end{equation}
Define
\begin{equation}
  \theta\equiv\phi-\delta,
  \qquad
  \Xi(\phi,\delta)\equiv\sin\phi-\sin(\phi-\delta).
  \label{eq:nonrelativistic-phase-functions}
\end{equation}
Then
\begin{align}
  n_x&=1+O(\beta_0^2),
  &
  n_z&=\frac{\beta_0}{\delta}\Xi+O(\beta_0^3),
  \label{eq:nonrelativistic-direction}\\
  b_r&=\beta_0\cos\theta+O(\beta_0^3),
  &
  a_r&=-\beta_0\sin\theta+O(\beta_0^3),
  \label{eq:nonrelativistic-retarded-kinematics}\\
  \kappa_r&=1+O(\beta_0^2),
  &
  \gamma_o&=1+O(\beta_0^2).
  \label{eq:nonrelativistic-kappa-gamma}
\end{align}
Substitution into Eqs.~\eqref{eq:dimensionless-local-Ex}--\eqref{eq:dimensionless-local-By}
yields
\begin{align}
  \frac{E_{\hat x}}{E_\omega}
  &=\frac{1}{\delta^2}+O(\beta_0^2),
  \label{eq:nonrelativistic-local-Ex}\\
  \frac{E_{\hat z}}{E_\omega}
  &=\beta_0\left[
  \frac{\Xi}{\delta^3}
  -\frac{\cos\theta}{\delta^2}
  +\frac{\sin\theta}{\delta}
  \right]+O(\beta_0^3),
  \label{eq:nonrelativistic-local-Ez}\\
  \frac{cB_{\hat y}}{E_\omega}
  &=\beta_0\left[
  \frac{\cos\theta-\cos\phi}{\delta^2}
  -\frac{\sin\theta}{\delta}
  \right]+O(\beta_0^3).
  \label{eq:nonrelativistic-local-By}
\end{align}
The leading transverse electric field is therefore precisely the Coulomb
field,
\begin{equation}
  E_{\hat x}
  =\frac{q}{4\pi\epsilon_0d^2}+O(\beta_0^2),
  \label{eq:nonrelativistic-coulomb-leading}
\end{equation}
whereas the longitudinal electric field and local magnetic field begin at
first order in the oscillation amplitude.

To second order in $\beta_0$, the proper-time and inertial-phase averages agree
for $S_{\hat x}$, because both the $1/\gamma_o$ weighting correction and the
correction to $T_\tau$ contribute only at $O(\beta_0^4)$.  Direct averaging of
Eqs.~\eqref{eq:nonrelativistic-local-Ez} and
\eqref{eq:nonrelativistic-local-By} gives
\begin{equation}
  \frac{\left\langle S_{\hat x}\right\rangle_\tau}{S_\omega}
  =\frac{\beta_0^2}{2\delta^3}
  \left(\delta-\sin\delta\right)
  +O(\beta_0^4).
  \label{eq:nonrelativistic-average-Sx}
\end{equation}
Since $\delta-\sin\delta>0$ for every $\delta>0$, the leading nonrelativistic
cycle-averaged transverse flux is directed from the source toward the observer.
Together with Eq.~\eqref{eq:average-Sz-zero}, this is an analytical answer to
the cycle-averaged local-flux question in the small-$\beta_0$ regime.

\subsection{Near-zone limit}
\label{subsec:near-zone-limit}

We next take $\delta\ll1$ at fixed $\beta_0<1$.  In addition to
$b_o=\beta_0\cos\phi$, define
\begin{equation}
  h_o\equiv\beta_0\sin\phi,
  \qquad
  \gamma_o=(1-b_o^2)^{-1/2}.
  \label{eq:near-zone-bo-ho}
\end{equation}
Expansion of the null condition gives
\begin{equation}
  \Delta
  =\gamma_o\delta
  +\frac{1}{2}b_oh_o\gamma_o^4\delta^2
  +O(\delta^3).
  \label{eq:near-zone-retardation-expansion}
\end{equation}
The leading factor $\gamma_o$ is the increased light-travel time caused by the
longitudinal motion during propagation across the transverse separation.

Expansion of the exact local fields produces
\begin{align}
  \frac{E_{\hat x}}{E_\omega}
  &=\frac{1}{\delta^2}
  -\frac{3}{8}\gamma_o^6h_o^2+O(\delta),
  \label{eq:near-zone-local-Ex}\\
  \frac{E_{\hat z}}{E_\omega}
  &=\frac{\gamma_o^3h_o}{2\delta}+O(1),
  \label{eq:near-zone-local-Ez}\\
  \frac{cB_{\hat y}}{E_\omega}
  &=\frac{1}{2}b_o\gamma_o^4
  \left(1-3\gamma_o^2h_o^2\right)+O(\delta).
  \label{eq:near-zone-local-By}
\end{align}
Several cancellations are evident.  The local transverse electric field has
no $\delta^{-1}$ correction: its leading term is the instantaneous transverse
Coulomb field.  The longitudinal field contains an induction-like
$d^{-1}$ contribution, while the magnetic field remains of order
$E_\omega/c$ as $\delta\to0$.  Nevertheless, the ideal point-charge energy
density is dominated by the Coulomb singularity,
\begin{equation}
  \frac{u_{\mathrm{loc}}}{u_\omega}
  =\frac{1}{2\delta^4}+O(\delta^{-2}).
  \label{eq:near-zone-energy-density}
\end{equation}
The limit $d\to0$ itself is excluded because the observer then coincides with
the singular source worldline.

In the overlap region $\beta_0\ll1$ and $\delta\ll1$,
Eq.~\eqref{eq:nonrelativistic-average-Sx} gives
\begin{equation}
  \frac{\left\langle S_{\hat x}\right\rangle_\tau}{S_\omega}
  =\frac{\beta_0^2}{12}
  +O(\beta_0^2\delta^2,\beta_0^4),
  \label{eq:nonrelativistic-near-average-Sx}
\end{equation}
because $\delta-\sin\delta=\delta^3/6+O(\delta^5)$.  This finite leading
mean flux is parametrically small compared with the divergent quasistatic
energy density.

\subsection{Far-zone limit}
\label{subsec:far-zone-limit}

For $\delta\gg1$ at fixed $\beta_0<1$, the bounded longitudinal displacement
implies
\begin{align}
  \Delta&=\delta+O(\delta^{-1}),
  &
  \phi_r&=\phi-\delta+O(\delta^{-1}),
  \label{eq:far-zone-retarded-geometry}\\
  n_x&=1+O(\delta^{-2}),
  &
  n_z&=O(\delta^{-1}),
  \qquad
  \kappa_r=1+O(\delta^{-1}).
  \label{eq:far-zone-direction-kappa}
\end{align}
Let
\begin{equation}
  a_\infty(\phi,\delta)
  \equiv-\beta_0\sin(\phi-\delta)
  \label{eq:far-zone-acceleration-phase}
\end{equation}
denote the leading retarded dimensionless acceleration.  The $R^{-1}$ terms
then dominate the local fields:
\begin{align}
  \frac{E_{\hat x}}{E_\omega}
  &=-\frac{\gamma_ob_oa_\infty}{\delta}
  +O(\delta^{-2}),
  \label{eq:far-zone-local-Ex}\\
  \frac{E_{\hat z}}{E_\omega}
  &=-\frac{a_\infty}{\delta}
  +O(\delta^{-2}),
  \label{eq:far-zone-local-Ez}\\
  \frac{cB_{\hat y}}{E_\omega}
  &=\frac{\gamma_oa_\infty}{\delta}
  +O(\delta^{-2}).
  \label{eq:far-zone-local-By}
\end{align}
The additional $E_{\hat x}$ component is generated by boosting the transverse
inertial-frame radiation field along $z$.  These leading fields satisfy
\begin{equation}
  \bm E_{\mathrm{loc}}\cdot\bm B_{\mathrm{loc}}=0,
  \qquad
  E_{\mathrm{loc}}^2=c^2B_{\mathrm{loc}}^2,
  \label{eq:far-zone-null-field}
\end{equation}
as expected for a null far-zone field.

The corresponding leading energy density and flux are
\begin{align}
  \frac{u_{\mathrm{loc}}}{u_\omega}
  &=\frac{\gamma_o^2a_\infty^2}{\delta^2}
  +O(\delta^{-3}),
  \label{eq:far-zone-energy-density}\\
  \frac{S_{\hat x}}{S_\omega}
  &=\frac{\gamma_oa_\infty^2}{\delta^2}
  +O(\delta^{-3}),
  \label{eq:far-zone-Sx}\\
  \frac{S_{\hat z}}{S_\omega}
  &=-\frac{\gamma_o^2b_oa_\infty^2}{\delta^2}
  +O(\delta^{-3}).
  \label{eq:far-zone-Sz}
\end{align}
Thus the leading energy-flow direction in the observer's local frame is
\begin{equation}
  \frac{\bm S_{\mathrm{loc}}}{cu_{\mathrm{loc}}}
  =\frac{1}{\gamma_o}\hat{\bm e}_{\hat x}
  -b_o\hat{\bm e}_{\hat z}
  +O(\delta^{-1}),
  \label{eq:far-zone-energy-flow-direction}
\end{equation}
whose leading magnitude is unity because
$\gamma_o^{-2}+b_o^2=1$.  Equation~\eqref{eq:far-zone-energy-flow-direction}
is precisely the aberration of a wave propagating predominantly along $+x$ in
$S$ into the observer's instantaneous rest frame.

The leading proper-time-averaged transverse flux can be evaluated in closed
form.  Combining Eqs.~\eqref{eq:proper-time-average-phase} and
\eqref{eq:far-zone-Sx} gives
\begin{equation}
  \frac{\left\langle S_{\hat x}\right\rangle_\tau}{S_\omega}
  =\frac{\pi\beta_0^2}
  {4E_{\mathrm{ell}}(\beta_0^2)\delta^2}
  +O(\delta^{-3}),
  \label{eq:far-zone-average-Sx}
\end{equation}
which is strictly positive.  In deriving this result, the factor $\gamma_o$ in
$S_{\hat x}$ cancels the $1/\gamma_o$ proper-time measure, leaving
$\int_0^{2\pi}a_\infty^2d\phi=\pi\beta_0^2$.  The longitudinal average remains
zero exactly by Eq.~\eqref{eq:average-Sz-zero}.

\subsection{Combined nonrelativistic far-zone limit}
\label{subsec:nonrelativistic-far-zone-limit}

For $\beta_0\ll1$ and $\delta\gg1$,
$E_{\mathrm{ell}}(\beta_0^2)=\pi/2+O(\beta_0^2)$, so
Eq.~\eqref{eq:far-zone-average-Sx} reduces to
\begin{equation}
  \frac{\left\langle S_{\hat x}\right\rangle_\tau}{S_\omega}
  =\frac{\beta_0^2}{2\delta^2}
  +O\left(\frac{\beta_0^4}{\delta^2},\frac{\beta_0^2}{\delta^3}\right).
  \label{eq:nonrelativistic-far-average-Sx-dimensionless}
\end{equation}
Restoring dimensions with
$\beta_0=A\omega/c$, $\delta=\omega d/c$, and
$S_\omega=\epsilon_0cE_\omega^2$ gives
\begin{equation}
  \left\langle S_{\hat x}\right\rangle_\tau
  =\frac{q^2A^2\omega^4}
  {32\pi^2\epsilon_0c^3d^2}
  \left[1+O(\beta_0^2)+O(\delta^{-1})\right].
  \label{eq:standard-dipole-flux-recovery}
\end{equation}
This is the standard cycle-averaged radiation flux of a nonrelativistic
oscillating charge observed in the direction transverse to its acceleration
\cite{Jackson1999,Rohrlich2007}.
Its recovery provides a normalization and sign check on the retarded geometry,
the tetrad projection, and the proper-time averaging prescription.

The near-zone, far-zone, and nonrelativistic limits emphasize different pieces
of the exact field and should not be interchanged without tracking their
assumptions.  In particular, $\delta\to0$ approaches the point-charge
singularity, while $\delta\to\infty$ isolates the null acceleration field.
The overlap formulas above show explicitly where the limits agree and provide
benchmarks for the numerical results that follow.

\begin{figure*}[t]
  \centering
  \includegraphics[width=0.9\textwidth]{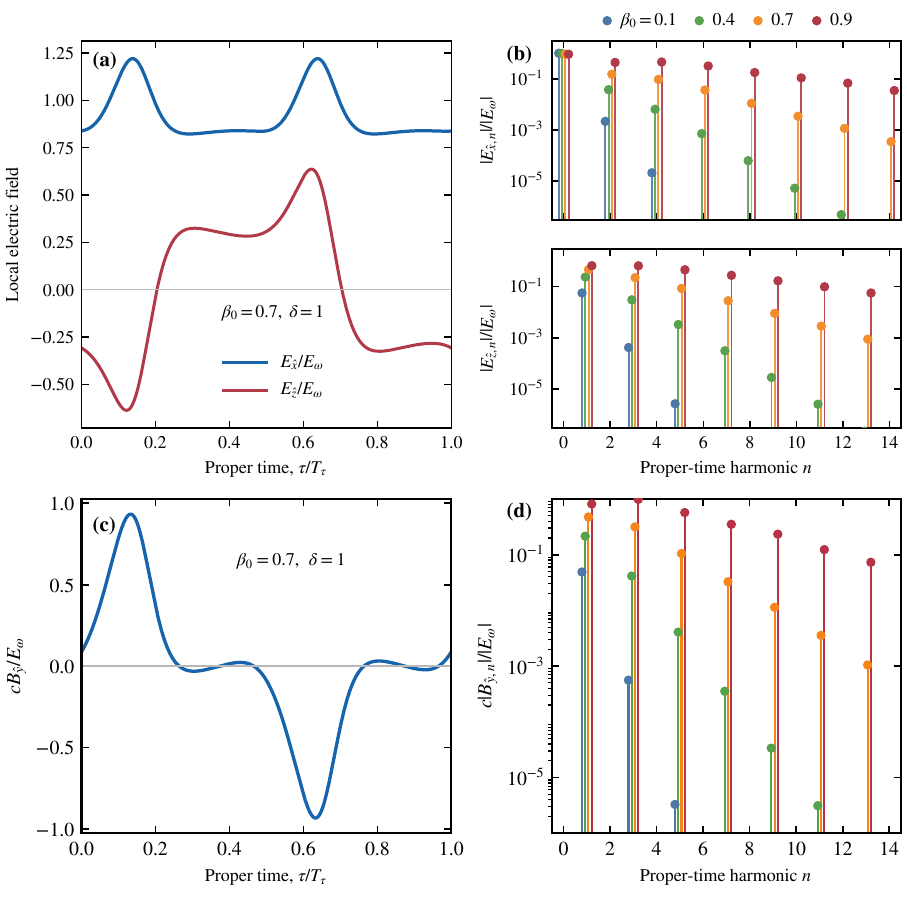}
  \caption{Proper-time waveforms and harmonic spectra of the locally measured
electric and magnetic fields at $\delta=1$.
(a) Exact dimensionless electric fields $E_{\hat x}/E_\omega$ and
$E_{\hat z}/E_\omega$ for $\beta_0=0.7$, plotted over one observer
proper-time period. The transverse component repeats after half a period,
whereas the longitudinal component changes sign.
(b) One-sided electric-field harmonic amplitudes for
$\beta_0=0.1$, $0.4$, $0.7$, and $0.9$. The upper and lower plots show,
respectively, the symmetry-allowed even harmonics of $E_{\hat x}$ and
odd harmonics of $E_{\hat z}$.
(c) Exact dimensionless magnetic field $cB_{\hat y}/E_\omega$ for
$\beta_0=0.7$, which changes sign after half a proper-time period.
(d) One-sided magnetic-field harmonic amplitudes
$c|B_{\hat y,n}|/|E_\omega|$ for the same four values of $\beta_0$,
with colors defined in (b). Only odd harmonics are allowed.
Each vertical stem represents one discrete harmonic; the four
$\beta_0$ series are displaced slightly horizontally for visibility.
Increasing $\beta_0$ produces stronger waveform distortion and a
progressively broader harmonic spectrum. All results are obtained
from the exact retarded-time solution and sampled uniformly in the observer's proper time.}
  \label{fig:proper-time-fields-spectra}
\end{figure*}

\section{Proper-time spectra}
\label{sec:proper-time-spectra}

The local fields derived in
Sec.~\ref{sec:observer-measured-field} are periodic functions
of the observer's proper time. Relativistic time dilation,
retardation, and the nonlinear dependence of the
Li\'enard--Wiechert field on the source velocity distort
the measured waveforms and generate higher harmonics.
We evaluate these effects numerically using the exact
retarded solution, with no expansion in $\beta_0$ or $\delta$.

Define the dimensionless local field components
\begin{equation}
  e_{\hat i}(\tau)\equiv
  \frac{E_{\hat i}(\tau)}{E_\omega},
  \qquad
  b_{\hat y}(\tau)\equiv
  \frac{cB_{\hat y}(\tau)}{E_\omega}.
  \label{eq:proper-time-field-normalization}
\end{equation}
The proper-time Fourier coefficients of the electric
components are
\begin{equation}
  \begin{aligned}
  c_{\hat i,n}
  &\equiv\frac{1}{T_\tau}
  \int_0^{T_\tau}e_{\hat i}(\tau)
  e^{-in\Omega_\tau\tau}\,d\tau,\\
  \Omega_\tau&\equiv\frac{2\pi}{T_\tau},
  \end{aligned}
  \label{eq:proper-time-fourier-coefficients}
\end{equation}
and those of the magnetic component are
\begin{equation}
  c^{(B)}_{\hat y,n}
  \equiv\frac{1}{T_\tau}
  \int_0^{T_\tau}b_{\hat y}(\tau)
  e^{-in\Omega_\tau\tau}\,d\tau.
  \label{eq:proper-time-magnetic-fourier-coefficients}
\end{equation}
Because the fields are real,
$c_{\hat i,-n}=c_{\hat i,n}^*$ and
$c^{(B)}_{\hat y,-n}=(c^{(B)}_{\hat y,n})^*$.
We display the one-sided harmonic amplitudes
\begin{equation}
  \begin{aligned}
  \mathcal A_{\hat i,0}&=|c_{\hat i,0}|,\\
  \mathcal A_{\hat i,n}&=2|c_{\hat i,n}|
  \quad(n\geq1),
  \end{aligned}
  \label{eq:one-sided-harmonic-amplitudes}
\end{equation}
and
\begin{equation}
  \begin{aligned}
  \mathcal A^{(B)}_{\hat y,0}&=|c^{(B)}_{\hat y,0}|,\\
  \mathcal A^{(B)}_{\hat y,n}&=2|c^{(B)}_{\hat y,n}|
  \quad(n\geq1).
  \end{aligned}
  \label{eq:one-sided-magnetic-harmonic-amplitudes}
\end{equation}
For $n\geq1$, these quantities give the amplitudes of
the corresponding real sinusoidal harmonics.
In Fig.~\ref{fig:proper-time-fields-spectra}, the
dimensional one-sided harmonic amplitudes are denoted
by $|E_{\hat i,n}|$ and $|B_{\hat y,n}|$, so that
\begin{equation}
  \begin{aligned}
  \mathcal A_{\hat i,n}
  &=\frac{|E_{\hat i,n}|}{|E_\omega|},\\
  \mathcal A^{(B)}_{\hat y,n}
  &=\frac{c|B_{\hat y,n}|}{|E_\omega|}.
  \end{aligned}
  \label{eq:dimensional-harmonic-amplitude-normalization}
\end{equation}
The factor $c$ makes the magnetic normalization
dimensionless in SI units.

The exact half-period relations already impose selection
rules before any numerical calculation. From
Eqs.~\eqref{eq:local-Ex-half-period}--
\eqref{eq:local-By-half-period},
\begin{align}
  c_{\hat x,n}&=0 &&\text{for odd }n,
  \label{eq:Ex-spectral-selection-rule}\\
  c_{\hat z,n}&=0 &&\text{for even }n,
  \label{eq:Ez-spectral-selection-rule}\\
  c^{(B)}_{\hat y,n}&=0 &&\text{for even }n.
  \label{eq:By-spectral-selection-rule}
\end{align}
Thus $E_{\hat x}$ may contain a dc component, whereas
$E_{\hat z}$ and $B_{\hat y}$ have zero proper-time mean.
The numerical spectra provide direct tests of these
symmetry predictions and quantify the allowed higher
harmonics.

For each $\beta_0$, the retarded interval $\Delta(\phi)$
is obtained from Eq.~\eqref{eq:retarded-root-function}
on a uniform inertial-phase grid, with the root solve
restricted to the analytical bracket
Eq.~\eqref{eq:retardation-bounds}. The exact local fields
are then evaluated from
Eqs.~\eqref{eq:dimensionless-local-Ex}--
\eqref{eq:dimensionless-local-By}.
We construct the monotonic proper-time map from
Eq.~\eqref{eq:proper-time-phase-map}, interpolate the
fields onto a uniform grid in $\tau/T_\tau$, and apply
a discrete Fourier transform over one complete period.
The calculations in
Fig.~\ref{fig:proper-time-fields-spectra} use $32769$
inertial-phase points and $16384$ uniform proper-time
samples. The endpoint at $\phi=2\pi$ is included for
quadrature and interpolation but excluded from the
Fourier grid. No window function is applied because
the record contains exactly one complete period.

Figure~\ref{fig:proper-time-fields-spectra}(a) shows the
exact electric fields at the representative parameters
$\beta_0=0.7$ and $\delta=1$. The waveforms are visibly
nonsinusoidal. The transverse field remains predominantly
positive and contains a nonzero dc component, while the
longitudinal field changes sign between the two half
cycles. The transverse component repeats after half a
proper-time period, whereas the longitudinal component
is antisymmetric under the same translation.
The numerical waveforms satisfy
\begin{align}
  \max_\tau
  \left|E_{\hat x}(\tau+T_\tau/2)-E_{\hat x}(\tau)\right|
  &<4\times10^{-14}|E_\omega|,
  \label{eq:numerical-Ex-half-period-error}\\
  \max_\tau
  \left|E_{\hat z}(\tau+T_\tau/2)+E_{\hat z}(\tau)\right|
  &<7\times10^{-14}|E_\omega|,
  \label{eq:numerical-Ez-half-period-error}
\end{align}
for this parameter set. These relations account for
the even and odd harmonic selection rules for the
transverse and longitudinal electric fields,
respectively.

Figure~\ref{fig:proper-time-fields-spectra}(b) shows the
electric-field harmonic spectra for $\beta_0=0.1$,
$0.4$, $0.7$, and $0.9$ at $\delta=1$.
The upper and lower plots display the symmetry-allowed
even harmonics of $E_{\hat x}$ and odd harmonics of
$E_{\hat z}$, respectively.
At $\beta_0=0.1$, the transverse spectrum is dominated
by its dc component and second harmonic, while the
longitudinal spectrum is dominated by its fundamental;
higher allowed harmonics decrease rapidly.
At $\beta_0=0.7$, allowed components remain visible
through the largest displayed harmonic.
At $\beta_0=0.9$, the higher harmonics become substantial:
for example,
$\mathcal A_{\hat x,14}\simeq3.5\times10^{-2}$ and
$\mathcal A_{\hat z,13}\simeq5.6\times10^{-2}$.
The electric spectra therefore show progressively
stronger departures from their nonrelativistic
harmonic content.

Figure~\ref{fig:proper-time-fields-spectra}(c) shows the
exact magnetic field for $\beta_0=0.7$ and $\delta=1$,
over the same proper-time period as the electric
waveforms. Its waveform is also visibly nonsinusoidal,
with pronounced positive and negative peaks separated
by half a period. Like the longitudinal electric field,
the magnetic field changes sign under a half-period
translation and therefore has zero proper-time mean.
Its instantaneous value is nevertheless generally
nonzero. The numerical waveform satisfies
\begin{equation}
  \max_\tau
  \left|b_{\hat y}(\tau+T_\tau/2)+b_{\hat y}(\tau)\right|
  <8\times10^{-14}.
  \label{eq:numerical-By-half-period-error}
\end{equation}
This antisymmetry requires the magnetic spectrum to
contain only odd harmonics.

Figure~\ref{fig:proper-time-fields-spectra}(d) shows the
magnetic-field harmonic spectra for $\beta_0=0.1$,
$0.4$, $0.7$, and $0.9$, using the color convention
of the electric spectra. Only odd harmonics are allowed,
and the dc component vanishes.
At $\beta_0=0.1$, the spectrum is dominated by its
fundamental, with higher odd harmonics decreasing
rapidly. Increasing $\beta_0$ strengthens the higher
harmonics and broadens the spectrum, as for the electric
fields. At $\beta_0=0.7$, magnetic harmonics remain
visible through $n=13$. At $\beta_0=0.9$, their amplitudes include $\mathcal A^{(B)}_{\hat y,1}\simeq0.812$, $\mathcal A^{(B)}_{\hat y,3}\simeq0.983$, and $\mathcal A^{(B)}_{\hat y,13}\simeq7.30\times10^{-2}$. The third magnetic harmonic exceeds the fundamental
for this parameter set, illustrating the substantial
distortion of the magnetic waveform at relativistic
oscillation speeds.

The broadening of both electric and magnetic spectra
arises from three coupled effects: the nonlinear
retarded-phase relation $\phi_r(\phi)$, relativistic
factors such as $\gamma_o$ and $\kappa_r^{-3}$, and the
nonlinear conversion from inertial phase to proper time.
The forbidden harmonics provide sensitive numerical
error measures. For all four parameter sets, the largest
forbidden electric amplitude through $n=14$ is less
than $5\times10^{-14}$ of the largest allowed amplitude.
The largest even magnetic amplitude through $n=14$,
including the dc component, is below $4\times10^{-14}$
in the dimensionless normalization used here.
Halving both the inertial-phase and proper-time grids
changes every displayed electric amplitude by less than
$2\times10^{-7}$ and every displayed magnetic amplitude
by less than $6\times10^{-9}$.
These tests verify Fourier convergence and the exact
symmetry selection rules for both fields.
The retarded-root, local field-transformation, and
field-invariant checks provide additional independent
validation.

The proper-time spectrum is an operational property
of this specified observer family. It should not be
identified directly with the spectrum in inertial
coordinate time or with the response of a detector
having finite bandwidth or memory
\cite{Mashhoon1990,Glauber1963}.
For the ideal local observer considered here, the
proper-time Fourier decomposition gives the natural
frequency content of the measured electric and magnetic
fields. The increasingly strong higher harmonics are
therefore relativistic features of the local observables,
rather than numerical artifacts. Their presence alone,
however, does not determine whether the complete
finite-distance field should be classified as radiation.

\section{Complementary non-inertial coordinate description}
\label{sec:noninertial-coordinates}

The preceding calculation was deliberately carried out in the inertial frame
$S$ and only afterward projected onto the observer's tetrad.  It is also useful
to describe the same physics in coordinates in which both the charge and the
observer are spatially at rest.  This complementary description makes clear
why being ``static'' in a non-inertial coordinate system does not, by itself,
imply a time-independent electromagnetic field or a vanishing locally measured
Poynting vector.

Introduce comoving coordinates $(cT,X,Y,Z)$ by the time-dependent translation
\begin{equation}
  T=t,
  \qquad
  X=x,
  \qquad
  Y=y,
  \qquad
  Z=z-A\sin(\omega T).
  \label{eq:comoving-coordinate-transformation}
\end{equation}
The source and observer worldlines then become
\begin{align}
  (X_q,Y_q,Z_q)&=(0,0,0),
  \label{eq:charge-static-comoving}\\
  (X_o,Y_o,Z_o)&=(d,0,0),
  \label{eq:observer-static-comoving}
\end{align}
respectively.  Thus their coordinate separation is the constant transverse
vector $d\,\hat{\bm X}$.  This transformation is a global change of coordinates on Minkowski spacetime, and it introduces no spacetime curvature. Writing
\begin{equation}
  \beta(T)=\beta_0\cos(\omega T),
  \label{eq:comoving-beta}
\end{equation}
the Minkowski line element becomes
\begin{align}
  ds^2={}&-c^2\bigl[1-\beta^2(T)\bigr]dT^2
  +2c\beta(T)\,dT\,dZ \nonumber\\
  &+dX^2+dY^2+dZ^2 .
  \label{eq:comoving-line-element}
\end{align}
In the coordinate order $(cT,X,Y,Z)$, the nonzero metric components are
$g_{00}=-\bigl(1-\beta^2\bigr)$, $g_{03}=g_{30}=\beta$, $g_{11}=g_{22}=g_{33}=1$, and $\det(g_{ab})=-1$.  The off-diagonal component $g_{03}$ is the shift
associated with the moving spatial coordinates.  The metric is explicitly
time dependent whenever the common velocity changes, and the coordinate-time
direction is not orthogonal to the hypersurfaces $T=\mathrm{const}$.
These coordinates inherit the simultaneity convention of $S$; they are not a
unique global rest frame constructed by Einstein synchronization, nor are
they Fermi normal coordinates away from a selected worldline.  They are,
however, sufficient for describing the present pair of worldlines and their
fixed transverse separation.

A worldline at fixed $(X,Y,Z)$ has
\begin{equation}
  d\tau=\sqrt{1-\beta^2(T)}\,dT,
  \label{eq:comoving-static-proper-time}
\end{equation}
in agreement with Eq.~\eqref{eq:proper-time-increment}.  The unit timelike
tetrad vector of such an observer has comoving-coordinate components
\begin{equation}
  e_{\hat 0}{}^a=\gamma(1,0,0,0).
  \label{eq:comoving-tetrad-time}
\end{equation}
The spatial coordinate vector $\partial_Z$ is not orthogonal to
$e_{\hat 0}$.  The orthonormal tetrad corresponding to
Eqs.~\eqref{eq:observer-tetrad-time}--\eqref{eq:observer-tetrad-z} is instead
\begin{align}
  e_{\hat 1}{}^a&=(0,1,0,0),
  &
  e_{\hat 2}{}^a&=(0,0,1,0),
  \label{eq:comoving-tetrad-transverse}\\
  e_{\hat 3}{}^a&=\left(\gamma\beta,0,0,\frac{1}{\gamma}\right).
  \label{eq:comoving-tetrad-longitudinal}
\end{align}
Direct contraction gives
$g_{ab}e_{\hat\alpha}{}^ae_{\hat\beta}{}^b
=\eta_{\hat\alpha\hat\beta}$.  Consequently, projecting the field tensor on
this tetrad reproduces exactly the local fields derived in Sec.~\ref{sec:observer-measured-field}.  In
particular, coordinate components such as $F_{0i}$ are not by themselves the
electric field measured by the static comoving observer, because the
coordinate basis is neither orthogonal nor normalized.

Maxwell's equations retain their generally covariant form
\cite{MisnerThorneWheeler1973},
\begin{align}
  \frac{1}{\sqrt{-g}}\partial_a
  \left(\sqrt{-g}\,F^{ab}\right)&=\mu_0J^b,
  \label{eq:maxwell-comoving-inhomogeneous}\\
  \partial_{[a}F_{bc]}&=0.
  \label{eq:maxwell-comoving-homogeneous}
\end{align}
Here $\sqrt{-g}=1$, but index raising involves the time-dependent metric in
Eq.~\eqref{eq:comoving-line-element}.  Moreover, the connection encoded
by this metric enters equivalent formulations written in terms of electric
and magnetic coordinate components.  The point charge has only a temporal
current component in these coordinates,
\begin{equation}
  J^a
  =cq\,\delta(X)\delta(Y)\delta(Z)(1,0,0,0),
  \label{eq:static-current-comoving}
\end{equation}
so the source is indeed coordinate static.  Nevertheless, the differential
operator governing its field is time dependent.  A coordinate-static source
therefore need not generate a coordinate-static solution.

The same conclusion follows directly from causality.  In the comoving chart,
the source event $(T_r,0,0,0)$ and observation event $(T,d,0,0)$ have a fixed
coordinate separation, but the null condition between them remains
\begin{equation}
  c^2(T-T_r)^2
  =d^2+A^2\left[\sin(\omega T)-\sin(\omega T_r)\right]^2,
  \label{eq:comoving-null-condition}
\end{equation}
which is identical to Eq.~\eqref{eq:retarded-time}.  The field at the
observer is thus controlled by the source kinematics at $T_r$, while the
observer's tetrad is fixed by its kinematics at $T$.  The retarded mismatch
that produced the nonzero $B_{\hat y}$ in Sec.~\ref{sec:observer-measured-field} has not disappeared; it has
been transferred from the coordinate motion of the worldlines into the
time-dependent geometry of the coordinate system.

Finally, the coordinate energy current should not be confused with the local
energy flux measured by the observer.  The latter is the tetrad projection of
the stress--energy tensor,
\begin{equation}
  u_{\mathrm{loc}}=T^{\hat0\hat0},
  \qquad
  S_{\hat i}=cT^{\hat0\hat i}.
  \label{eq:comoving-local-stress-energy-projection}
\end{equation}
These projections are coordinate independent and reproduce
Eqs.~\eqref{eq:local-energy-density}--\eqref{eq:local-Sz}.  By contrast, a
coordinate component such as $T^0{}_i$ depends on the nonorthogonal basis and
does not directly represent the reading of a local energy-flux detector.

Although Maxwell's equations can be solved directly in the comoving
coordinates, doing so provides no practical simplification for the present
problem.  The source is spatially static, but the metric is explicitly time
dependent and nonorthogonal, so the coordinate Maxwell equations contain
time-dependent couplings among the field components.  Retardation must still
be enforced through the null relation between the source and observation
events.  Furthermore, the resulting coordinate components must ultimately be
projected onto an orthonormal tetrad before they can be identified with fields
measured by the observer.

The inertial-frame projection method used here is therefore the more direct
route to the physical observables.  One first evaluates the standard retarded
Li\'enard--Wiechert tensor in $S$ and then performs the algebraic tetrad
projection in Eq.~\eqref{eq:field-tensor-tetrad-projection} at the observation
event.  This procedure is not an approximation.  Tensor covariance guarantees
that it is exactly equivalent to solving Maxwell's equations in the
non-inertial coordinates and then projecting the resulting tensor onto the
same observer tetrad.  Its advantage is computational rather than conceptual:
it cleanly separates the retarded-field calculation from the
observer-dependent measurement operation.

The non-inertial description therefore agrees fully with the inertial-frame
calculation.  The source and observer are spatially static, but the comoving
metric, the retarded null relation, and the observer's orthonormal frame remain
time dependent.  The time-dependent electric field, nonzero magnetic field,
local energy flux, and proper-time harmonics found above are consequently not
coordinate artifacts.  They are local tensor observables evaluated along the
specified accelerated worldline.

\section{Discussion}
\label{sec:discussion}

\subsection{Relation to previous observer constructions}
\label{subsec:previous-observers}

The literature on accelerated charges contains statements that can appear
contradictory when the observer, spacetime region, and radiation criterion are
not specified.  The present calculation does not resolve those statements by
selecting one universal definition of radiation.  Instead, it fixes a concrete
source worldline, a concrete observer worldline, and an operational local
measurement prescription.  Within that specification, the electric field,
magnetic field, energy density, and Poynting vector are unambiguous.

The classic uniformly accelerated problem differs from the present one in
several essential respects.  For eternal hyperbolic motion, uniformly
coaccelerating observers occupy a Rindler wedge bounded by a horizon.  The
Born field can be stationary and purely electric within the accessible wedge,
whereas inertial observers can associate radiative energy with a different
spacetime region \cite{Born1909,FultonRohrlich1960,Rohrlich1963,Boulware1980,
EriksenGron2004,AlmeidaSaa2006}.  Our charge instead undergoes bounded harmonic
motion, its acceleration changes sign, and the prescribed observer follows a
parallel worldline at fixed transverse displacement.  There is no permanent
Rindler horizon separating this observer from an asymptotic radiative region.
The horizon-based stationary-field argument for eternal uniform acceleration
therefore does not apply directly.

Our observer also differs from an observer selected specifically to make a
given electromagnetic field static or purely electric.  Leonov showed that,
subject to stated conditions, one may construct special observer families for
which the field of an arbitrarily moving charge has such properties
\cite{Leonov2012}.  Here the observer is fixed independently of the field: its
worldline is chosen to share the source's velocity and acceleration at equal
inertial time while remaining transversely separated from it.  This kinematic
condition does not imply the field-adapted condition
Eq.~\eqref{eq:local-magnetic-zero-condition}.  The observer therefore measures
a generally nonzero $B_{\hat y}$ because the field at $t$ depends on the
source at $t_r$, whereas the local tetrad is determined by the observer at
$t$.

More general analyses of radiation in accelerated laboratories emphasize the
relative motion between a source and an observer congruence
\cite{FugmannKretzschmar1991}.  Our example gives a particularly transparent
form of the relevant distinction.  The source and observer have identical
kinematics at simultaneous events in $S$, but the emission and observation
events connected by a light ray are not simultaneous.  Equal instantaneous
velocity and acceleration therefore do not eliminate their retarded
kinematic mismatch.  Likewise, operational discussions of a comoving
detector \cite{Broda2012} concern what a specified apparatus can register,
which is logically distinct from asking whether an asymptotic radiative flux
exists.  The present tetrad fields supply the local electromagnetic input for
such a detector model, but do not replace the model itself.

These comparisons show that our result is not in conflict with earlier
observer constructions.  Different conclusions arise because different
worldlines, accessible regions, and measurement criteria define different
physical questions.  A transformation to coordinates in which the charge is
spatially static is not sufficient to identify our observer with a Rindler
observer or with a field-adapted observer.  What matters for a local
measurement is the electromagnetic tensor at the observation event and the
orthonormal tetrad carried by the specified observer.

\subsection{Local flux and the meaning of radiation}
\label{subsec:radiation-interpretation}

The word ``radiation'' is used for several related but inequivalent
properties.  It is therefore useful to separate four statements that can be
made about the present system.

First, the comoving observer measures time-dependent local fields.  Except in
special limits or at isolated phases, both $E_{\hat z}$ and $B_{\hat y}$ are
nonzero, and the local Poynting vector obtained in Sec.~\ref{sec:local-energy-flux} is generally
nonzero.  These are pointwise operational statements.  They do not depend on
a coordinate decomposition of the field and are not removed by making the
source and observer coordinate static.

Second, the retarded Li\'enard--Wiechert field contains a term proportional to
$R^{-1}$ as well as the velocity term proportional to $R^{-2}$.  The former is
conventionally called the acceleration or radiation field.  This separation
is especially useful asymptotically, where the $R^{-1}$ term dominates the
field and its stress--energy scales as $R^{-2}$.  At finite distance, however,
the total local energy density and flux also contain the velocity field and
interference between the velocity and acceleration contributions.  Assigning
the complete finite-distance Poynting vector to radiation would therefore mix
radiative transport with bound and induction-zone energy flow.  Covariant
bound--radiative decompositions can sharpen this distinction, but they add a
criterion beyond the pointwise field measurement \cite{Teitelboim1970}.

Third, the far-zone result has the standard properties of radiation.  The
leading field is null, Eq.~\eqref{eq:far-zone-null-field}, its energy propagates
locally at speed $c$, and its cycle-averaged transverse flux falls as $d^{-2}$.
In the combined nonrelativistic and far-zone limit,
Eq.~\eqref{eq:standard-dipole-flux-recovery} reproduces the familiar radiation
flux of an oscillating point charge.  Thus the system radiates in the usual
inertial asymptotic sense.  The comoving observer at large transverse
separation measures the Lorentz-transformed version of that same null field,
including the aberrated energy-flow direction in
Eq.~\eqref{eq:far-zone-energy-flow-direction}.

Fourth, detection by a physical apparatus is a further operational question.
An ideal pointlike observer, under the hypothesis of locality, assigns the
instantaneous fields calculated here.  Whether a real detector absorbs net
energy or registers a particular spectrum depends on its coupling,
orientation, bandwidth, spatial extent, and memory.  The proper-time Fourier
coefficients in Sec.~\ref{sec:proper-time-spectra} are therefore the spectrum of the local driving
field, not by themselves a prediction of detector counts or absorbed power.
A response function must be supplied before the latter quantities can be
computed.

These distinctions lead to a calibrated interpretation.  It is correct to
say that the prescribed comoving observer measures a time-dependent
electromagnetic field, a generally nonzero magnetic field, and a local energy
flux.  It is also correct to identify the leading far-zone contribution as
radiation.  At finite $d$, however, a nonzero instantaneous or cycle-averaged
local Poynting vector is not, by itself, sufficient to classify the entire
field as radiation.  Indeed, the exact invariant in
Eq.~\eqref{eq:local-field-invariant-scalar} is positive at every finite
retarded distance, whereas a pure radiation field is null.  The exact field
therefore retains a nonradiative component even when its far-zone part is
unambiguously radiative.

This interpretation also separates local flux from source energy loss.  A
complete energy-balance statement would require integrating an appropriately
defined flux over a closed boundary and accounting for the work performed by
the external agent that sustains the charge's motion, together with bound-field
energy and radiation reaction.  The value of $\bm S_{\mathrm{loc}}$ on one
observer worldline cannot by itself provide that global balance.  Our result
answers what this observer measures; it does not infer the total radiated
power from a single local sample.

\subsection{A photoemission thought experiment and open questions}
\label{subsec:photoemission-thought-experiment}

The original motivation for the present calculation may be formulated as a
photoemission thought experiment that is not restricted to harmonic motion.
The experiment invokes the conversion of incident electromagnetic energy into
discrete electron-emission events \cite{Einstein1905}.
More generally, consider any prescribed timelike trajectory $z(t)$ and the
pair of worldlines
\begin{equation}
  \bm r_q(t)=(0,0,z(t)),
  \qquad
  \bm r_o(t)=(d,0,z(t)).
  \label{eq:general-photoemission-worldlines}
\end{equation}
Replace the ideal pointlike observer by a small photocathode carried along the
second worldline.  In an inertial frame, the retarded electromagnetic field
produced by the accelerated charge can reach the photocathode at a later
observation event.  Suppose that the detector records $N$ emitted
photoelectrons during a specified interval of its proper time.  The
integer-valued detector record $N$ is an operationally invariant result: a
change from inertial to non-inertial coordinates cannot make the recorded
events disappear.  The corresponding detection rate should naturally be
defined with respect to the detector proper time, $dN/d\tau$, rather than an
arbitrary coordinate time.  The sinusoidal worldlines studied in this paper
are one explicit realization of Eq.~\eqref{eq:general-photoemission-worldlines}.

For the sinusoidal realization, in the comoving coordinates of Sec.~\ref{sec:noninertial-coordinates}, both
the charge and the photocathode are spatially static.  Nevertheless, the
metric is time dependent and
nonorthogonal, the retarded null relation remains nontrivial, and the
photocathode is exposed to the time-dependent local fields derived in Sec.~\ref{sec:observer-measured-field}.
There is therefore no contradiction between coordinate-static worldlines and
a nonzero detector response.  Our classical calculation determines the local
electromagnetic field that would drive such a detector, but it does not
determine the number of emitted photoelectrons.

The thought experiment raises several questions that lie beyond the present
work.  Is the photoemission probability governed directly by the proper-time
spectrum obtained in Sec.~\ref{sec:proper-time-spectra}?  Can the bound and induction parts of the
finite-distance field produce photoelectrons, and, if so, can such a response
be operationally distinguished from absorption of asymptotic radiation?  How
should the photoelectric threshold be formulated for an accelerated detector
with finite size and response time?  Would a semiclassical description of
quantum matter coupled to the classical local field be sufficient, or is a
quantized electromagnetic field required?  In quantum photodetection theory,
such probabilities depend on a specified detector--field coupling and on field
correlation functions, illustrating why the local classical spectrum alone is
not a complete count-rate prediction \cite{Glauber1963}.

Uniform proper acceleration poses an especially sharp version of this
thought experiment.  A charge and a transversely separated coaccelerating
detector may both be stationary in Rindler coordinates, while inertial
observers assign radiation to the accelerated charge.  If an explicit
detector model predicts photoemission for the bounded motion studied here,
does it also predict a response for an eternally uniformly accelerated charge?
Alternatively, can the Rindler horizon and the stationarity of the field in
the accessible wedge prevent the coaccelerating detector from accessing the
radiative part of the field \cite{Boulware1980,AlmeidaSaa2006}?  Does the
answer change when uniform acceleration is maintained for only a finite
interval, so that no eternal acceleration horizon is present?  The thought
experiment applies to all these trajectories, but its outcome need not be
independent of the acceleration history and causal structure.

A related question concerns energy accounting.  If photoelectrons are
produced, how should the absorbed energy be divided among radiative energy
from the charge, reversible bound-field energy, and work performed by the
external agents that maintain the prescribed motions of the source and
detector?  Answering this question requires an explicit detector model,
backaction, and a global energy-balance analysis.  We leave these issues open.
The present results provide the exact local classical field and its
proper-time spectrum as inputs for such a future study.

\subsection{Scope and limitations}
\label{subsec:scope-limitations}

The present model was chosen to isolate retardation and local measurement in
the simplest nontrivial comoving geometry.  The source is an ideal point
charge executing prescribed sinusoidal motion for all times, and the observer
is idealized as pointlike.  The calculation excludes the singular source
worldline by requiring $d>0$.  It also neglects radiation reaction and the
backaction of a detector; an unspecified external force is implicitly assumed
to maintain the prescribed harmonic trajectory.

The observer's tetrad is Fermi--Walker transported, but the translated
coordinate system of Sec.~\ref{sec:noninertial-coordinates} should not be interpreted as a unique global
proper reference frame for an extended rigid laboratory.  It inherits the
simultaneity convention of $S$, and the congruence of coordinate-static
worldlines need not constitute a globally Born-rigid system in the
longitudinal direction.  For the two worldlines used here, the separation is
purely transverse, so their fixed distance $d$ and common proper-time
synchronization are nevertheless well defined by the construction in
Sec.~\ref{sec:configuration}.

Our use of the hypothesis of locality neglects nonlocal acceleration effects
that may arise for a detector whose dimensions or response time are not small
compared with the acceleration and field-variation scales.  Likewise, the
proper-time spectrum assumes an ideal sampling of the local classical field
over complete periods.  Finite-time measurements would introduce spectral
windowing, while a quantum detector would require a separate interaction
Hamiltonian and field-state specification.

Several extensions follow naturally.  One may treat unequal source and
observer motions, include longitudinal as well as transverse separation,
consider a finite acceleration history rather than eternal periodic motion,
or replace the point observer by a specified detector model.  A global
stress--energy analysis could compare the local flux with radiated power and
bound-field energy on extended hypersurfaces.  Finally, the same
inertial-solution-plus-tetrad-projection strategy applies to more general
accelerated trajectories and, locally, to electromagnetic measurements in
curved spacetime whenever an appropriate field solution is available.

\section{Conclusions}
\label{sec:conclusions}

We have obtained the exact retarded electromagnetic field of a harmonically
oscillating point charge as measured by an observer undergoing the same motion
at a fixed transverse separation.  Although the source and observer have
identical velocities and accelerations at every common inertial time, this
simultaneous comotion does not make the locally measured field static.  The
field at the observation event is determined by the charge at an earlier
retarded event, while the observer's tetrad is determined by its motion at the
observation event.  This retarded kinematic mismatch is the central physical
origin of the result.

Starting from the Li\'enard--Wiechert field in an inertial frame, we projected
the electromagnetic tensor onto a Fermi--Walker-transported tetrad carried by
the observer.  The resulting expressions are exact for every
$\beta_0=A\omega/c<1$ and every nonzero
$\delta=\omega d/c$.  The observer measures two generally time-dependent
electric components, $E_{\hat x}$ and $E_{\hat z}$, together with a generally
nonzero magnetic component $B_{\hat y}$.  Thus equality of the simultaneous
source and observer velocities is not sufficient to transform away the local
magnetic field.

The projected stress--energy tensor gives an unambiguous local energy density
and Poynting vector.  The instantaneous energy flux generally has both
transverse and longitudinal components.  Symmetry requires the proper-time
average of the longitudinal component to vanish exactly.  The transverse
average is directed from the source toward the observer in the analytically
established nonrelativistic regime and is strictly positive at leading order
in the far zone.  In the combined nonrelativistic and far-zone limit, it
reduces to the standard cycle-averaged radiation flux of an oscillating point
charge observed transverse to its acceleration.

The analytical limits clarify the different physical contributions to the
exact field.  In the near zone, the transverse electric field is dominated by
the instantaneous Coulomb term, while retardation produces induction-like and
magnetic corrections.  In the far zone, the acceleration field dominates,
the leading electromagnetic field becomes null, and the locally measured
energy flow is the aberrated version of the conventional inertial radiation
field.  At finite distance, however, the exact field remains non-null and
contains velocity, induction, and interference contributions.  A nonzero
local Poynting vector alone therefore does not identify the entire
finite-distance field as radiation.

The proper-time analysis reveals an additional observable consequence of the
accelerated motion.  Relativistic time dilation, nonlinear retardation, and
the velocity dependence of the Li\'enard--Wiechert field generate
nonsinusoidal local waveforms and higher harmonics.  Exact half-period
symmetries require $E_{\hat x}$ to contain only even proper-time harmonics,
including a possible dc component, whereas $E_{\hat z}$ and $B_{\hat y}$
contain only odd harmonics.  The numerical spectra verify these selection
rules and show a progressive broadening as $\beta_0$ approaches unity.

We also formulated the problem in coordinates in which both worldlines are
spatially static.  In that description the metric is time dependent and
nonorthogonal, and the retarded null relation remains unchanged.  Directly
solving Maxwell's equations in those coordinates is possible but offers no
practical simplification.  Evaluating the standard inertial retarded field and
then projecting it onto the observer's tetrad is exactly equivalent by tensor
covariance and provides a substantially more direct route to the measurable
quantities.

The present work therefore answers the local operational question without
requiring a universal definition of radiation: the specified transversely
comoving observer measures a time-dependent electromagnetic field, a
generally nonzero magnetic field, and a generally nonzero local energy flux.
The far-zone part is unambiguously radiative, whereas interpretation of the
complete finite-distance flux requires a distinction among radiative, bound,
and induction-field energy.  Extending the calculation to finite-duration
motion, unequal source and observer trajectories, and detectors with explicit
response functions would provide natural next steps toward connecting these
local classical observables with experimentally registered signals. Its connection to electromagnetic radiation in a gravitational field remains a long-standing open problem.

\appendix

\section{Li\'enard--Wiechert conventions and tensor components}
\label{app:lw-conventions}

This appendix records the sign, index, and unit conventions used throughout
the calculation \cite{Jackson1999,MisnerThorneWheeler1973}.  In inertial
coordinates $x^\mu=(ct,x,y,z)$, the metric is
$\eta_{\mu\nu}=\operatorname{diag}(-1,1,1,1)$ and
$\epsilon^{123}=+1$.  The Lorentz force and field tensor are defined by
\begin{equation}
  \frac{dp^\mu}{d\tau}=qF^{\mu\nu}u_\nu,
  \qquad
  F^{0i}=\frac{E_i}{c},
  \qquad
  F^{ij}=\epsilon^{ijk}B_k.
  \label{eq:app-field-conventions}
\end{equation}
Lowering both indices gives $F_{0i}=-E_i/c$ and $F_{ij}=F^{ij}$.
For the planar geometry of this work, the contravariant and covariant tensors
are therefore
\begin{align}
  F^{\mu\nu}
  &={}
  \begin{pmatrix}
    0&E_x/c&0&E_z/c\\
    -E_x/c&0&0&-B_y\\
    0&0&0&0\\
    -E_z/c&B_y&0&0
  \end{pmatrix},
  \label{eq:app-field-tensor-contravariant}\\
  F_{\mu\nu}
  &={}
  \begin{pmatrix}
    0&-E_x/c&0&-E_z/c\\
    E_x/c&0&0&-B_y\\
    0&0&0&0\\
    E_z/c&B_y&0&0
  \end{pmatrix}.
  \label{eq:app-field-tensor-covariant}
\end{align}

The retarded source quantities appearing in the Li\'enard--Wiechert solution
are evaluated before any differentiation is performed.  Thus
$\dot{\bm\beta}_r$ means $d\bm\beta/dt_r$ evaluated at the retarded event,
not a derivative with respect to the observation time.  With
$X^\mu=x_o^\mu(t)-x_q^\mu(t_r)$, the retarded event satisfies
$X^\mu X_\mu=0$ and $X^0>0$.  The scalar
\begin{equation}
  \rho_r\equiv-\frac{X_\mu u_r^\mu}{c}
  =\gamma_rR\kappa_r>0
  \label{eq:app-invariant-retarded-distance}
\end{equation}
is the invariant retarded distance.  The standard three-vector field in
Eq.~\eqref{eq:lienard-wiechert-electric} is written instead in terms of
$R\kappa_r=\rho_r/\gamma_r$.

The electromagnetic stress--energy tensor is
\begin{equation}
  T^{\mu\nu}=\frac{1}{\mu_0}
  \left(F^{\mu\alpha}F^\nu{}_{\alpha}
  -\frac{1}{4}\eta^{\mu\nu}F^{\alpha\beta}F_{\alpha\beta}\right).
  \label{eq:app-stress-energy}
\end{equation}
In any orthonormal frame with the same signature,
\begin{equation}
  T^{\hat0\hat0}=u_{\mathrm{loc}},
  \qquad
  cT^{\hat0\hat i}=S_{\hat i}.
  \label{eq:app-stress-energy-observables}
\end{equation}
These relations fix the signs used for the local Poynting components in
Eqs.~\eqref{eq:local-Sx}--\eqref{eq:local-Sz}.

Two invariant identities are especially useful for checking all component
calculations.  With the conventions above,
\begin{align}
  \frac{1}{2}F_{\mu\nu}F^{\mu\nu}
  &=B^2-\frac{E^2}{c^2},
  \label{eq:app-first-invariant}\\
  \bm E\cdot\bm B&=0
  \label{eq:app-second-invariant}
\end{align}
for the present planar field.  Both identities are unchanged by the tetrad
projection.

\section{Explicit field components}
\label{app:explicit-components}

For direct implementation it is convenient to eliminate the intermediate
quantities $n_x$, $n_z$, $b_r$, and $a_r$.  Define
\begin{equation}
  \theta\equiv\phi-\Delta=\phi_r,
  \qquad
  \chi\equiv\sin\phi-\sin\theta,
  \qquad
  K\equiv1-\frac{\beta_0^2\chi\cos\theta}{\Delta},
  \label{eq:app-explicit-abbreviations}
\end{equation}
where $\Delta$ is the unique positive solution of
Eq.~\eqref{eq:retarded-root-function}.  The null condition implies
\begin{equation}
  \Delta^2=\delta^2+\beta_0^2\chi^2,
  \qquad K=\kappa_r.
  \label{eq:app-explicit-null-condition}
\end{equation}

The complete inertial-frame components can then be written as
\begin{align}
  \frac{E_x}{E_\omega}
  &={}
  \frac{\delta}{K^3\Delta^3}
  \left[1-\beta_0^2\cos^2\theta
  -\beta_0^2\chi\sin\theta\right],
  \label{eq:app-explicit-inertial-Ex}\\
  \frac{E_z}{E_\omega}
  &={}
  \frac{\beta_0}{K^3\Delta^3}
  \left\{
  \left[1-\beta_0^2\cos^2\theta\right]
  \left(\chi-\Delta\cos\theta\right)
  +\delta^2\sin\theta
  \right\},
  \label{eq:app-explicit-inertial-Ez}\\
  \frac{cB_y}{E_\omega}
  &={}
  \frac{\beta_0\delta}{K^3\Delta^3}
  \left\{
  \left[1-\beta_0^2\cos^2\theta\right]\cos\theta
  -\Delta\sin\theta
  \right\}.
  \label{eq:app-explicit-inertial-By}
\end{align}
These expressions are algebraically equivalent to
Eqs.~\eqref{eq:dimensionless-Ex}--\eqref{eq:dimensionless-By}; no expansion
has been made.

For the local fields, also define
\begin{align}
  b_o&\equiv\beta_0\cos\phi,
 \label{eq:B11}\\
  \gamma_o&=(1-b_o^2)^{-1/2},
 \label{eq:B12}\\
  b_r&\equiv\beta_0\cos\theta,
 \label{eq:B13}\\
  a_r&\equiv-\beta_0\sin\theta.
  \label{eq:app-explicit-observer-source-kinematics}
\end{align}
Substitution into Eqs.~\eqref{eq:dimensionless-local-Ex}--
\eqref{eq:dimensionless-local-By} gives
\begin{align}
  \frac{E_{\hat x}}{E_\omega}
  &={}
  \frac{\gamma_o\delta}{K^3\Delta^3}
  \left\{
  (1-b_r^2)(1-b_ob_r)
  +a_r(\beta_0\chi-b_o\Delta)
  \right\},
  \label{eq:app-explicit-local-Ex}\\
  \frac{E_{\hat z}}{E_\omega}
  &={}
  \frac{1}{K^3\Delta^3}
  \left\{
  (1-b_r^2)(\beta_0\chi-b_r\Delta)
  -a_r\delta^2
  \right\},
  \label{eq:app-explicit-local-Ez}\\
  \frac{cB_{\hat y}}{E_\omega}
  &={}
  \frac{\gamma_o\delta}{K^3\Delta^3}
  \left\{
  (1-b_r^2)(b_r-b_o)
  +a_r(\Delta-b_o\beta_0\chi)
  \right\}.
  \label{eq:app-explicit-local-By}
\end{align}
The remaining local components vanish.  Equations
\eqref{eq:app-explicit-local-Ex}--\eqref{eq:app-explicit-local-By}, followed by
Eqs.~\eqref{eq:dimensionless-local-energy-density}--
\eqref{eq:dimensionless-local-Sz}, form a compact implementation of all local
observables from the single scalar root $\Delta(\phi)$.

As elementary checks, setting $\beta_0=0$ gives
$E_{\hat x}/E_\omega=\delta^{-2}$ and all other field components zero.
Under $\phi\mapsto\phi+\pi$, uniqueness gives
$\theta\mapsto\theta+\pi$ and leaves $\Delta$ and $K$ invariant.  The explicit
formulas then immediately yield the parity relations in
Eqs.~\eqref{eq:local-Ex-half-period}--\eqref{eq:local-By-half-period}.

\section{Tetrad and Fermi--Walker transport}
\label{app:tetrad-transport}

The one-dimensional nature of the motion makes the observer tetrad most
transparent in terms of the rapidity $\zeta(\tau)$
\cite{MisnerThorneWheeler1973,MalufFaria2008}, defined by
\begin{equation}
  \tanh\zeta=b_o,
  \qquad
  \cosh\zeta=\gamma_o,
  \qquad
  \sinh\zeta=\gamma_ob_o.
  \label{eq:app-rapidity}
\end{equation}
The timelike and longitudinal tetrad vectors are
\begin{align}
  e_{\hat0}{}^\mu&=(\cosh\zeta,0,0,\sinh\zeta),
  \label{eq:app-rapidity-timelike}\\
  e_{\hat3}{}^\mu&=(\sinh\zeta,0,0,\cosh\zeta),
  \label{eq:app-rapidity-longitudinal}
\end{align}
while $e_{\hat1}{}^\mu=(0,1,0,0)$ and
$e_{\hat2}{}^\mu=(0,0,1,0)$.  Orthonormality follows directly from
$\cosh^2\zeta-\sinh^2\zeta=1$.

Since $db_o/dt=a_{o,z}/c$ and $dt/d\tau=\gamma_o$,
\begin{equation}
  \frac{d\zeta}{d\tau}
  =\gamma_o^3\frac{a_{o,z}}{c}
  =\frac{\alpha_o}{c}.
  \label{eq:app-rapidity-rate}
\end{equation}
Differentiating Eqs.~\eqref{eq:app-rapidity-timelike} and
\eqref{eq:app-rapidity-longitudinal} therefore gives
\begin{equation}
  \frac{de_{\hat0}{}^\mu}{d\tau}
  =\frac{\alpha_o}{c}e_{\hat3}{}^\mu,
  \qquad
  \frac{de_{\hat3}{}^\mu}{d\tau}
  =\frac{\alpha_o}{c}e_{\hat0}{}^\mu,
  \label{eq:app-tetrad-derivatives}
\end{equation}
with vanishing derivatives for the two transverse axes.  Using
$u_o^\mu=ce_{\hat0}{}^\mu$ and
$a_o^\mu=\alpha_oe_{\hat3}{}^\mu$ verifies the Fermi--Walker equation
\begin{equation}
  \frac{DV^\mu}{d\tau}
  =\frac{1}{c^2}
  \left(u_o^\mu a_{o\nu}-a_o^\mu u_{o\nu}\right)V^\nu
  \label{eq:app-fermi-walker-general}
\end{equation}
for every tetrad vector $V^\mu=e_{\hat\alpha}{}^\mu$.  Because all boosts are
collinear, there is no Thomas--Wigner rotation.

For completeness, direct tensor projection gives
\begin{align}
  -cF_{\hat0\hat1}
  &=-cF_{\mu\nu}e_{\hat0}{}^\mu e_{\hat1}{}^\nu
  =\gamma_o(E_x-b_ocB_y),
  \label{eq:app-project-Ex}\\
  -cF_{\hat0\hat3}
  &=E_z,
  \label{eq:app-project-Ez}\\
  F_{\hat3\hat1}
  &=\gamma_o\left(B_y-\frac{b_o}{c}E_x\right).
  \label{eq:app-project-By}
\end{align}
Since $B_{\hat y}=F_{\hat3\hat1}$ with the orientation adopted here, these
relations reproduce Eqs.~\eqref{eq:local-Ex-transformation}--
\eqref{eq:local-By-transformation}.  They also show explicitly that the
projection is a local Lorentz boost, not a transformation evaluated at the
retarded source event.

\section{Numerical retarded-time algorithm}
\label{app:numerical-algorithm}

The numerical calculation requires only a one-dimensional root solve at each
observation phase.  For fixed $(\phi,\beta_0,\delta)$, define $g(\Delta;\phi)$
as in Eq.~\eqref{eq:retarded-root-function}.  Its derivative is
\begin{equation}
  \frac{\partial g}{\partial\Delta}
  =2\left\{
  \Delta-\beta_0^2
  \left[\sin\phi-\sin(\phi-\Delta)\right]
  \cos(\phi-\Delta)
  \right\}.
  \label{eq:app-retarded-root-derivative}
\end{equation}
The analytical interval
\begin{equation}
  \Delta_- =\delta,
  \qquad
  \Delta_+=\sqrt{\delta^2+4\beta_0^2}
  \label{eq:app-root-bracket}
\end{equation}
brackets the unique physical root because
$g(\Delta_-)\leq0$ and $g(\Delta_+)\geq0$.

We use a safeguarded Newton iteration.  After evaluating $g$ at the current
iterate, the sign of $g$ updates the corresponding side of the bracket.  The
Newton proposal
\begin{equation}
  \Delta_{k+1}^{(N)}
  =\Delta_k-
  \frac{g(\Delta_k;\phi)}
  {\partial g(\Delta_k;\phi)/\partial\Delta}
  \label{eq:app-newton-step}
\end{equation}
is accepted only if it is finite and lies inside the updated bracket;
otherwise the bracket midpoint is used.  Iteration stops only when both the
step size and the null residual meet their prescribed tolerances.  The
previous phase's solution may be used as the initial value because
$\Delta(\phi)$ is smooth and periodic, although this continuation is not
required for convergence.

After finding the root, the calculation proceeds as follows:
\begin{enumerate}
  \item Set $\phi_r=\phi-\Delta$ and evaluate
  $(\chi,n_x,n_z,b_r,a_r,\kappa_r)$.
  \item Compute the exact local fields from
  Eqs.~\eqref{eq:dimensionless-local-Ex}--
  \eqref{eq:dimensionless-local-By}, or equivalently from Appendix~B.
  \item Construct the monotonic proper-time map by cumulative quadrature of
  $d\tau/d\phi=1/(\omega\gamma_o)$.
  \item Interpolate each field onto a uniform grid
  $\tau_j/T_\tau=j/N_\tau$, $j=0,\ldots,N_\tau-1$.
  \item Apply a real discrete Fourier transform and normalize the coefficients
  according to Eq.~\eqref{eq:one-sided-harmonic-amplitudes}.
\end{enumerate}
The endpoint at $\phi=2\pi$ is included in the phase grid for quadrature and
interpolation but excluded from the uniform Fourier grid so that the periodic
sample is not duplicated.  No window is applied because the record contains
exactly one complete proper-time period.

Several independent diagnostics are imposed.  At every phase we check
\begin{equation}
  |g|<\varepsilon_g,
  \qquad
  |n_x^2+n_z^2-1|<\varepsilon_n,
  \qquad
  \kappa_r>0,
  \label{eq:app-pointwise-numerical-checks}
\end{equation}
and compare the numerical derivative of $\phi_r(\phi)$ with
$\kappa_o/\kappa_r$.  On the proper-time grid we test the exact half-period
relations and the forbidden Fourier parities.  Finally, the entire calculation
is repeated after reducing the phase and proper-time grids; changes in the
reported amplitudes provide the convergence estimate quoted in Sec.~\ref{sec:proper-time-spectra}.

For Fig.~\ref{fig:proper-time-fields-spectra}, the phase grid contains
$32769$ points, the Fourier grid contains $16384$ points, and the root residual
is below $2\times10^{-15}$.  The symmetry and forbidden-parity tests are more
sensitive than the plotted curves and provide useful end-to-end checks of the
root solve, interpolation, and Fourier normalization.

\section{Asymptotic expansions}
\label{app:asymptotic-expansions}

This appendix supplies intermediate steps for the limits summarized in
Sec.~\ref{sec:analytical-limits}.  The expansions are made at fixed observation phase, and the order of
limits is kept explicit.

\subsection{Small-\texorpdfstring{$\beta_0$}{beta0} expansion}
\label{app:small-beta-expansion}

At fixed $\delta>0$, write $\Delta=\delta+\beta_0^2D_2+O(\beta_0^4)$.
Insertion into the null condition gives
\begin{equation}
  D_2=\frac{1}{2\delta}
  \left[\sin\phi-\sin(\phi-\delta)\right]^2,
  \label{eq:app-small-beta-D2}
\end{equation}
which reproduces Eq.~\eqref{eq:nonrelativistic-retardation-expansion}.
Because the correction to $\Delta$ starts at second order, the quantities
needed through first order in the fields are
\begin{align}
  n_x&=1+O(\beta_0^2), 
  \label{eq:E11}\\
  n_z&=\frac{\beta_0}{\delta}
  \left[\sin\phi-\sin(\phi-\delta)\right]+O(\beta_0^3),
  \label{eq:app-small-beta-direction}\\
  b_r&=\beta_0\cos(\phi-\delta)+O(\beta_0^3),
  \label{eq:E12}\\
  a_r&=-\beta_0\sin(\phi-\delta)+O(\beta_0^3),
  \label{eq:app-small-beta-kinematics}\\
  \kappa_r&=1+O(\beta_0^2),
  \label{eq:E13}\\
  \gamma_o&=1+O(\beta_0^2).
  \label{eq:app-small-beta-denominators}
\end{align}
Substitution produces Eqs.~\eqref{eq:nonrelativistic-local-Ex}--
\eqref{eq:nonrelativistic-local-By}.  Products of the first-order
$E_{\hat z}$ and $B_{\hat y}$ terms give the leading averaged transverse flux
in Eq.~\eqref{eq:nonrelativistic-average-Sx}.

\subsection{Small-\texorpdfstring{$\delta$}{delta} expansion}
\label{app:small-delta-expansion}

Let $b_o=\beta_0\cos\phi$, $h_o=\beta_0\sin\phi$, and seek
$\Delta=D_1\delta+D_2\delta^2+O(\delta^3)$.  Expanding
\begin{equation}
  \beta_0\left[\sin\phi-\sin(\phi-\Delta)\right]
  =b_o\Delta+\frac{h_o}{2}\Delta^2
  -\frac{b_o}{6}\Delta^3+O(\Delta^4)
  \label{eq:app-small-delta-chi}
\end{equation}
inside the null condition gives
\begin{equation}
  D_1=\gamma_o,
  \qquad
  D_2=\frac{1}{2}b_oh_o\gamma_o^4,
  \label{eq:app-small-delta-coefficients}
\end{equation}
and hence Eq.~\eqref{eq:near-zone-retardation-expansion}.  Useful associated
series are
\begin{align}
  b_r&=b_o+h_o\Delta-\frac{b_o}{2}\Delta^2+O(\Delta^3),
  \label{eq:app-small-delta-br}\\
  a_r&=-h_o+b_o\Delta+\frac{h_o}{2}\Delta^2+O(\Delta^3),
  \label{eq:app-small-delta-ar}\\
  n_z&=b_o+\frac{h_o}{2}\Delta-\frac{b_o}{6}\Delta^2
  +O(\Delta^3),
  \label{eq:app-small-delta-nz}\\
  n_x&=\frac{\delta}{\Delta}.
  \label{eq:app-small-delta-nx}
\end{align}
Although individual velocity and acceleration contributions contain several
inverse powers of $\delta$, their combination produces the cancellations
displayed in Eqs.~\eqref{eq:near-zone-local-Ex}--
\eqref{eq:near-zone-local-By}.  In particular, the coefficient of
$\delta^{-1}$ in $E_{\hat x}$ vanishes.

\subsection{Large-\texorpdfstring{$\delta$}{delta} expansion}
\label{app:large-delta-expansion}

Define the zeroth-order retarded-phase difference
\begin{equation}
  \chi_\infty\equiv\sin\phi-\sin(\phi-\delta).
  \label{eq:app-chi-infinity}
\end{equation}
At fixed $\beta_0<1$, iteration of the null condition gives
\begin{align}
  \Delta
  &=\delta+\frac{\beta_0^2\chi_\infty^2}{2\delta}
  +O(\delta^{-2}),
  \label{eq:app-large-delta-Delta}\\
  \phi_r
  &=\phi-\delta-\frac{\beta_0^2\chi_\infty^2}{2\delta}
  +O(\delta^{-2}),
  \label{eq:app-large-delta-phir}\\
  n_x&=1+O(\delta^{-2}),
 \label{eq:app-large-delta-x}\\
  n_z&=\frac{\beta_0\chi_\infty}{\delta}+O(\delta^{-2}),
  \label{eq:app-large-delta-direction}\\
  \kappa_r
  &=1-\frac{\beta_0^2\chi_\infty\cos(\phi-\delta)}{\delta}
  +O(\delta^{-2}).
  \label{eq:app-large-delta-kappa}
\end{align}
At leading order,
$a_r=-\beta_0\sin(\phi-\delta)+O(\delta^{-1})=a_\infty+O(\delta^{-1})$.
Retaining the $R^{-1}$ terms in the local fields gives
Eqs.~\eqref{eq:far-zone-local-Ex}--\eqref{eq:far-zone-local-By}.  Squaring
those terms gives the energy density and flux in
Eqs.~\eqref{eq:far-zone-energy-density}--\eqref{eq:far-zone-Sz}; subleading
velocity-field and interference contributions begin one inverse power of
$\delta$ later in the stress--energy tensor.

Finally, expanding the elliptic integral as
\begin{equation}
  E_{\mathrm{ell}}(\beta_0^2)
  =\frac{\pi}{2}
  \left(1-\frac{\beta_0^2}{4}+O(\beta_0^4)\right)
  \label{eq:app-elliptic-small-beta}
\end{equation}
in Eq.~\eqref{eq:far-zone-average-Sx} yields the combined
nonrelativistic far-zone expression
Eq.~\eqref{eq:nonrelativistic-far-average-Sx-dimensionless} and, after
restoring dimensions, Eq.~\eqref{eq:standard-dipole-flux-recovery}.

\bibliography{oscillating_charge_references}

\end{document}